\documentclass{ptephy_v1}%%%%%% to generate preprint number with ptep logo

\newtheorem{theorem}{Theorem}
\newtheorem{definition}{Definition}
\newtheorem{proposition}{Proposition}

\begin{document}

\title{Transversely trapping surfaces: Dynamical version}

%%%% To generate auto affiliation numbers please use \author{}\affil{} command

%<<<<<<<<<<<<< AFFILIATION >>>>>>>>>>>>>>>%

\author{Hirotaka Yoshino${}^1$}
\author{Keisuke Izumi${}^{2,3}$}
\author{Tetsuya Shiromizu${}^{3,2}$}
\author{Yoshimune Tomikawa${}^4$}
\affil{${}^1$Advanced Mathematical Institute, Osaka City University, Osaka 558-8585, Japan}
\affil{${}^2$Kobayashi-Maskawa Institute, Nagoya University, Nagoya 464-8602, Japan}
\affil{${}^3$Department of Mathematics, Nagoya University, Nagoya 464-8602, Japan}
\affil{${}^4$Faculty of Economics, Matsuyama University, Matsuyama 790-8578, Japan}

%%% To include the collaborator name... Please use the command "\collaborator"
%%% For example: \collaborator{ATLAS Collaboration}

\begin{abstract}
  We propose new concepts,
  {\it a dynamically transversely trapping surface
  (DTTS)} and {\it a marginally DTTS,}
  as indicators for a strong gravity region. 
  A DTTS is defined as a two-dimensional closed surface
  on a spacelike hypersurface
  such that photons emitted from arbitrary points on it
  in transverse directions are acceleratedly contracted
  in time, and a marginally DTTS
  is reduced to the photon sphere in 
  spherically symmetric cases.
  (Marginally) DTTSs have a close analogy with (marginally)
  trapped surfaces in many aspects. 
  After preparing the method of solving for a marginally DTTS
  in the time-symmetric initial data
  and the momentarily stationary axisymmetric initial data,
  some examples of marginally DTTSs are numerically constructed
  for systems of two black holes in the Brill--Lindquist initial
  data and in the Majumdar--Papapetrou spacetimes.
  Furthermore, the area of a DTTS
  is proved to satisfy the Penrose-like inequality,
  $A_0\le 4\pi (3GM)^2$, under some assumptions.
  Differences and connections between a DTTS
  and the other two concepts proposed by us previously,
  a loosely trapped surface 
  [Prog.\ Theor.\ Exp.\ Phys.\ {\bf 2017}, 033E01 (2017)]
  and a static/stationary
  transversely trapping surface 
  [Prog.\ Theor.\ Exp.\ Phys.\ {\bf 2017}, 063E01 (2017)],
  are also discussed. 
  A (marginally) DTTS provides us with a theoretical tool to significantly 
  advance our understanding of strong gravity fields.
  Also, since DTTSs are located outside the event horizon,
  they could possibly be related with future observations
  of strong gravity regions in dynamical evolutions.
  
\end{abstract}

\subjectindex{E0, E31, A13}

\maketitle

%======================================%
%<<<<<<<<<<<< SECTION I  >>>>>>>>>>>>>>%
%======================================%
%
\section{Introduction}
\label{section1}

There are two characteristic positions in a black hole spacetime.
Taking a Schwarzschild spacetime as an example, one is
the horizon $r=2GM$ that determines the black hole region,
where $r$ is the circumferential radius,
$G$ is the Newtonian constant of gravitation, and 
$M$ is the Arnowitt--Deser--Misner (ADM) mass that represents
the total gravitational energy evaluated at spatial infinity.
The region on and inside the horizon, $r\le 2GM$, is not observable
for distant observers. 
The other is the photon sphere $r=3GM$ on which circular
orbits of photons exist. The photon sphere is related to
various observable phenomena. For example, excitation
of quasinormal modes of fields is closely
related to the photon sphere \cite{Cardoso:2008},
and hence it affects the
gravitational waveform from, e.g., the merger of two black holes
(see Ref.~\cite{Abbott:2016}  for the first detection).
Also, the edge of the black hole shadow
in electromagnetic observations is
determined by the photon sphere \cite{Virbhadra:1999}, 
or its extension, the fundamental photon orbits \cite{Cunha:2017}.
Recently, the Event Horizon Telescope Collaboration
has succeeded in observing the black hole shadow of
a massive object at the center of the galaxy M87 \cite{EHTCollaboration:2019}.

There are various extended concepts of the horizon $r=2GM$,
and the most famous ones are an event horizon and
an apparent horizon. An event horizon is defined as
the outer boundary of a black hole region from which
nothing can escape to the outside region.
Specifying the position of the event horizon is crucial
to understanding the structure of a spacetime.
An apparent horizon is a two-dimensional surface
such that the expansion of outgoing null geodesics emitted 
from it vanishes.
Assuming cosmic censorship, the singularity theorem tells us that 
the presence of an apparent horizon implies the existence of
an event horizon outside,
and thus restricts the global properties
of the spacetime (see Chapt.~12.2 of Ref.~\cite{Wald}). 
Extended concepts of $r=2GM$ like the event and apparent horizons,
if properly defined, provide us with tools to greatly
advance our understanding of the properties of
spacetimes.  
This fact motivates us to consider extended concepts of the photon sphere
$r=3GM$.
In this paper, we study extended concepts
of a photon sphere to characterize 
a strong gravity region outside a black hole.

One of the extended concepts of a photon sphere 
is a photon surface proposed in Ref.~\cite{Claudel:2000}.
A photon surface is defined as a timelike surface
$S$ such that arbitrary photons emitted on arbitrary points on $S$
in arbitrary null tangent directions to $S$
continue to propagate on $S$.
Unlike a photon sphere, a photon surface
may change its shape and size in time.
For example, a hyperboloid in a flat Minkowski spacetime
becomes a photon surface \cite{Claudel:2000}. Therefore, the existence of
a photon surface does not necessarily imply the presence of
a strong gravity region. One may expect that
a static photon surface 
in a static spacetime would become an indicator for
the presence of a strong gravity region.
However, the definition of a photon surface imposes 
such a strong condition on the behavior of photons that 
the spacetime must be highly symmetric 
in order to possess a static photon surface.
Related to this, various uniqueness theorems
have been studied for spacetimes with photon surfaces
\cite{Cederbaum:2014,Cederbaum:2015a,Yazadjiev:2015a,Cederbaum:2015b,Yazadjiev:2015b,Yazadjiev:2015c,Rogatko:2016,Yoshino:2016,Tomikawa:2016,Tomikawa:2017}.
See also Refs.~\cite{Gibbons:2016,Koga:2016,Shoom:2017,Koga:2018,Koga:2019a,Koga:2019b}
for related discussions.

The present authors have also suggested two concepts to characterize
a strong gravity region: A loosely trapped surface (LTS) \cite{Shiromizu:2017}
and a transversely trapping surface (TTS) \cite{Yoshino:2017}
(referred to as a static/stationary TTS in this paper). 
An LTS is defined from the behavior of the mean curvature $k$ in a flow of
two-dimensional surfaces 
on a spacelike hypersurface $\Sigma$. 
A static/stationary TTS is defined as a static/stationary timelike surface
$S$ in a static/stationary spacetime
such that arbitrary photons emitted on arbitrary points on $S$
in arbitrary null tangent directions to $S$
propagate on $S$ or to the inside region of $S$.
Then, inspired by the Penrose inequality \cite{Penrose:1973} 
\begin{equation}
  A_{\rm AH}\le 4\pi (2GM)^2,
  \label{Penrose-inequality}
\end{equation}
which is conjectured (and partly proved)
to be satisfied by the area of an apparent horizon $A_{\rm AH}$, 
we have proved that the area $A_0$
of each of an LTS and the spatial section
of a static/stationary TTS satisfies a Penrose-like inequality,
\begin{equation}
  A_0\le 4\pi (3GM)^2,
  \label{Penroselike-inequality}
\end{equation}
under some assumptions (see also earlier work for a photon sphere
in Ref.~\cite{Hod:2017}). The concept of a
static/stationary TTS is used to classify surfaces
in a Kerr spacetime \cite{Galtsov:2019a,Galtsov:2019b}. 
Shortcomings of our previous works
are that in the case of an LTS, the relation to the behavior
of photons cannot be read from the definition directly,
and in the case of a static/stationary TTS,
its straightforward generalization to dynamical
cases does not necessarily represent
a strong gravity region similarly to a photon surface. 
We consider that it is necessary to introduce
a concept of a surface as a strong gravity indicator
defined from the photon behavior, 
with applicability to dynamically evolving spacetimes.

Motivated by the above discussions, the purpose of this paper
is threefold. First,
we define a new concept, {\it a dynamically transversely trapping surface
(DTTS),} that satisfies the above requirements. 
Roughly speaking, a DTTS is defined as a two-dimensional surface
on a spacelike hypersurface such that 
photons emitted in transverse directions experience
accelerated contraction due to strong gravity.
A marginally DTTS is defined as a special case. 
The concept of a (marginally) DTTS
has close analogy with a (marginally) trapped surface
(or an apparent horizon).
Note that the concept of a DTTS in this paper
is different from that of a static/stationary TTS in
Ref.~\cite{Yoshino:2017}.
This point will be discussed in detail.

Second,
we show that a (marginally) DTTS is a well-defined concept, 
by explicitly constructing some examples numerically.  
We prepare the method of solving for marginally DTTSs
in some restricted configurations, i.e. the time-symmetric
initial data and the momentarily stationary axisymmetric initial data.
Then, marginally DTTSs are numerically solved for 
in the systems of two equal-mass black holes in the
Brill--Lindquist initial data \cite{Brill:1963}
and the Majumdar--Papapetrou spacetime \cite{Majumdar:1947,Papapetrou:1947}.

Third,
we clarify some of the general properties of DTTSs.
To be specific, we prove that the area of a DTTS satisfies the
Penrose-like inequality in Eq.~\eqref{Penroselike-inequality}
under some assumptions. We also prove that
there are configurations where 
a DTTS is guaranteed to be an LTS at the same time.

This paper is organized as follows.
In Sect.~\ref{section2} we define the concepts of a DTTS,
a dynamically transversely trapping region, and
a marginally DTTS.
In Sect.~\ref{section3} we specify the configurations to be studied,
i.e. the time-symmetric initial data and the momentarily stationary
axisymmetric initial data, 
and useful formulas to
calculate (marginally) DTTSs are presented.
In Sect.~\ref{section4}, marginally DTTSs are explicitly constructed
numerically 
for two-black-hole systems in the Brill--Lindquist initial data.
In Sect.~\ref{section5}, marginally DTTSs are calculated
for systems of two extremal black holes in the Majumdar--Papapetrou
spacetimes. In Sect.~\ref{section6}
we prove that a DTTS satisfies
the Penrose-like inequality in Eq.~\eqref{Penroselike-inequality}
under some assumptions, and we discuss the connection between DTTSs
and LTSs in Sect.~\ref{section7}. 
Section~\ref{section8} is devoted to a summary and discussions.
In Appendices~\ref{Appendix-A} and \ref{Appendix-B}, detailed derivations of
the equations for marginally DTTSs in the Brill--Lindquist initial
data and in the Majumdar--Papapetrou spacetimes are presented, respectively.
Throughout the paper, 
we study in the framework of the theory of general relativity
for four-dimensional spacetimes.  
We use the units in which the speed of light is unity, $c=1$,
while the Newtonian constant of gravitation $G$ is explicitly shown.

%
%======================================%
%<<<<<<<<<<<< SECTION II  >>>>>>>>>>>>>>%
%======================================%
%
\section{Definition of dynamically transversely trapping surfaces}
\label{section2}

In this section we present the definition of a DTTS.
In Sect.~\ref{section2-1},
we examine photon surfaces in a Schwarzschild spacetime
in order to learn a lesson that motivates the definition of a DTTS.
Then, a DTTS is defined in Sect.~\ref{section2-2}, and its meaning
is discussed in Sect.~\ref{section2-3}.
In Sect.~\ref{section2-4}, analogies between DTTSs and 
trapped surfaces are discussed.

\subsection{Motivation from the Schwarzschild spacetime}
\label{section2-1}

Let us begin our discussion by examining photon surfaces
in a Schwarzschild spacetime. Since a photon surface
is composed of null geodesics, we study null geodesic equations
below. The metric of a Schwarzschild
spacetime is
\begin{equation}
  ds^2 = -f(r)dt^2+\frac{dr^2}{f(r)} + r^2(d\theta^2+\sin^2\theta d\phi^2),
\end{equation}
with
\begin{equation}
f(r):=1-\frac{2GM}{r}.
\end{equation}
From the $t$ and $\phi$ components of null geodesic equations
on the equatorial plane $\theta=\pi/2$, we obtain
the conservation laws of energy and angular momentum,
\begin{eqnarray}
  f(r)\dot{t} &=& E,
  \label{geodesic-energy-conservation}\\
  r^2\dot{\phi} &=& Eb,\label{geodesic-angular-momentum-conservation}
\end{eqnarray}
where dot denotes the derivative with the affine parameter $\lambda$
of the null geodesic, and $E$ and $b$ are the energy and the impact
parameter, respectively. The null condition leads to
\begin{equation}
  f(r)\dot{t}^2 = \frac{\dot{r}^2}{f(r)}+r^2\dot{\phi}^2.
  \label{geodesic-null-condition}
\end{equation}
From these equations, we have
\begin{equation}
  \frac{dr}{dt} = \pm f(r)\sqrt{1-\frac{b^2}{r^2}f(r)}.
  \label{geodesic-Schwarschild-drdt}
\end{equation}
Solutions of the radial coordinate
for the geodesic equations are obtained
by integrating this equation. If $b>3\sqrt{3}\ GM$, the geodesic
is confined in the region $r>3GM$ or $r<3GM$, and possesses 
a pericenter or an apocenter, respectively.
If $b<3\sqrt{3}\ GM$, the geodesic does not have a turning point
and crosses the photon sphere $r=3GM$
from the inside region (resp. outside region) to the outside region
(resp. inside region) according to the plus (resp. minus)
sign in Eq.~\eqref{geodesic-Schwarschild-drdt}
(see Ref.~\cite{Gralla:2019} for a recent study on black hole shadows
that has close connection to such behavior of photons).
For a later convenience, we present the radial geodesic equations,
\begin{equation}
  \ddot{r} = \frac{f^\prime}{2f}\dot{r}^2 -\frac{ff^\prime}{2}\dot{t}^2
  +fr\dot{\phi}^2 = \frac{E^2b^2}{r^3}\left(1-\frac{3GM}{r}\right),
  \label{geodesic-Schwarzschild-ddotr}
\end{equation}
where prime indicates derivative with respect
to $r$ and the second equality is derived using
Eqs.~\eqref{geodesic-energy-conservation}--\eqref{geodesic-null-condition}.

If a solution $r(t)$ to Eq.~\eqref{geodesic-Schwarschild-drdt}
is obtained, 
we can construct a photon surface \cite{Claudel:2000} 
by virtue of the spherical symmetry.
At $t=0$, we consider photons with the same values of the 
radial position $r$ and the radial velocity $dr/dt$ but with
arbitrary angular positions and arbitrary angular
velocities. Since all such photons have the same
radial dependence $r(t)$, they form a photon surface $S$.
Since $r(t)$ can be arbitrarily large,
the photon surface itself does not necessarily imply the
existence of a strong gravity region.
Therefore, we must find a quantity that is suitable as an indicator
for a strong gravity region.

The induced geometry of $S$ is given by
\begin{equation}
  ds^2 = -\alpha^2 dt^2 + r^2(d\theta^2+\sin^2\theta d\phi^2),
\end{equation}
with the lapse function
\begin{equation}
\alpha = \frac{b}{r}f(r).
\end{equation}
We denote $t=\mathrm{const.}$ surfaces in $S$ by $\sigma_t$,
and its induced metric and extrinsic curvature by $h_{ab}$
and $\bar{k}_{ab}:=(1/2){}^{(3)}\bar{\pounds}_{\bar{n}}h_{ab}$, respectively,
where ${}^{(3)}\bar{\pounds}$ and $\bar{n}^a$ denote the
Lie derivative in $S$ and the future-directed unit normal to
$\sigma_{t}$ in $S$.  
The trace of the extrinsic curvature is 
\begin{equation}
\bar{k} = \frac{2}{bf(r)}\frac{dr}{dt}.
\end{equation}
Using Eq.~\eqref{geodesic-Schwarzschild-ddotr},
the Lie derivative of $\bar{k}$ with respect to $\bar{n}^a$ is calculated as
\begin{equation}
  {}^{(3)}\bar{\pounds}_{\bar{n}}\bar{k}
  = \frac{2}{r^2}\left(1-\frac{3GM}{r}\right).
\end{equation}
Then, we find that if $\sigma_t$ is located outside (resp. inside)
the photon sphere $r=3GM$, the value of ${}^{(3)}\bar{\pounds}_{\bar{n}}\bar{k}$
is positive (resp. negative). Physically, a photon surface
is in accelerated expansion or decelerated contraction
in the region $r>3GM$, and in decelerated expansion or
accelerated contraction in the region $r<3GM$, reflecting
the strength of gravity. 
Therefore, ${}^{(3)}\bar{\pounds}_{\bar{n}}\bar{k}$
is a good indicator for a strong gravity region,
and this result motivates us to define a DTTS
using the quantity ${}^{(3)}\bar{\pounds}_{\bar{n}}\bar{k}$.

\subsection{Definition}
\label{section2-2}

%===========<FIGURE1>============%
%
\begin{figure}[tb]
\centering
\includegraphics[width=0.7\textwidth,bb=0 0 401 237]{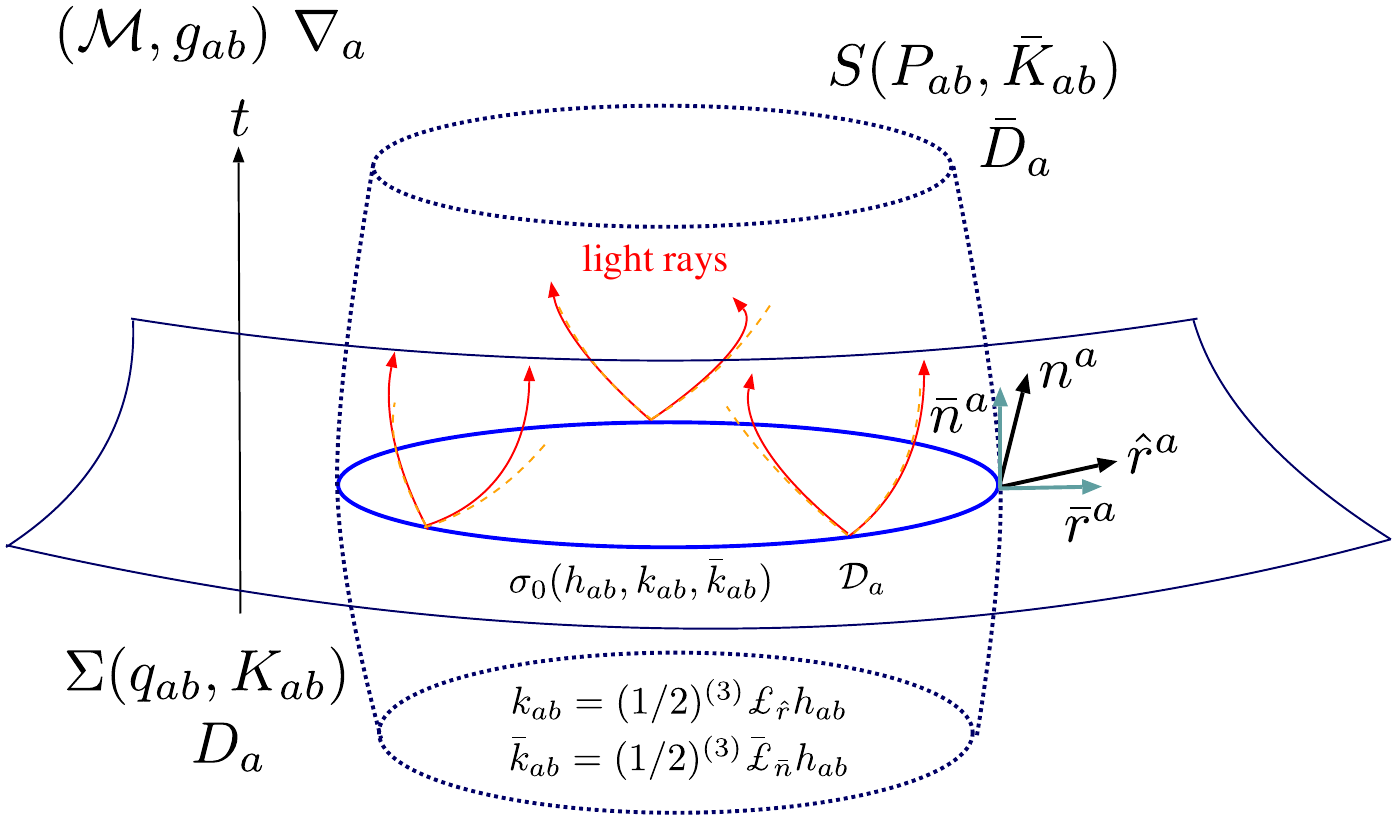}
\caption{
  Configuration to be considered.
  We consider a two-dimensional closed surface $\sigma_0$ in a
  spacelike hypersurface $\Sigma$ of a spacetime $\mathcal{M}$.
  A timelike hypersurface $S$ intersects with $\Sigma$
  precisely at $\sigma_0$. 
  Notations are also indicated. See text for details.
}
\label{schematic-DTTS}
\end{figure}
%
%=================================%

Before introducing the definition of a DTTS,
it is appropriate to show the configuration to be considered
and specify notations. Figure~\ref{schematic-DTTS}
presents the typical configuration to be studied.
We consider a spacetime $\mathcal{M}$ with the metric $g_{ab}$,
and $\nabla_a$ is the covariant derivative associated with $g_{ab}$.
In $\mathcal{M}$, take a spacelike hypersurface $\Sigma$ whose
future-directed timelike unit normal is $n^a$. A spatial metric
$q_{ab}$ is induced on $\Sigma$ as
\begin{equation}
  q_{ab}=g_{ab}+n_an_b,
\end{equation}and
the extrinsic curvature is defined as
\begin{equation}
K_{ab} = \frac12\pounds_n q_{ab},
\end{equation}
where $\pounds$ is a Lie derivative with respect to $\mathcal{M}$.
The derivative operator associated with $q_{ab}$ is denoted by $D_a$. 
In $\Sigma$, we take a two-dimensional closed orientable surface
$\sigma_0$ in order to examine whether it is a DTTS or not.
The induced metric of $\sigma_0$ is
\begin{equation}
  h_{ab}=q_{ab}-\hat{r}_a\hat{r}_b,
\end{equation}
where $\hat{r}^a$ is the spacelike unit normal to $\sigma_0$
in $\Sigma$ (namely, $\hat{r}^a$ is tangential to $\Sigma$).
The extrinsic curvature of $\sigma_0$ as a hypersurface
in $\Sigma$ is introduced as
\begin{equation}
  k_{ab} = \frac12{}^{(3)}\pounds_{\hat{r}}h_{ab},
  \label{def:kab}
\end{equation}
where ${}^{(3)}\pounds$ is a Lie derivative with respect to $\Sigma$.
The covariant derivative associated with $h_{ab}$ is denoted by $\mathcal{D}_a$.

We introduce a timelike hypersurface $S$ in $\mathcal{M}$, which
intersects with $\Sigma$ precisely at $\sigma_0$.
Note that the two hypersurfaces
$S$ and $\Sigma$ are not necessarily orthogonal to each other
at $\sigma_0$, and therefore, the outward spacelike unit normal $\bar{r}^a$
to $S$ does not agree with $\hat{r}^a$ in general.
The metric induced on $S$ is 
\begin{equation}
  \bar{p}_{ab}=g_{ab}-\bar{r}_a\bar{r}_b,
 \label{def:barpab}
\end{equation}
and the extrinsic curvature of $S$ is defined by
\begin{equation}
  \bar{K}_{ab} = \frac12\pounds_{\bar{r}}\bar{p}_{ab}.
  \label{def:barKab}
\end{equation}
The covariant derivative associated with $\bar{p}_{ab}$ is denoted by
$\bar{D}_a$. The two-dimensional surface $\sigma_0$ can be regarded
as a hypersurface in $S$, and the future directed unit normal
to $\sigma_0$ in this sense is denoted by $\bar{n}^a$.
Note that $\bar{n}^a$
and $n^a$ are different from each other in general.
The extrinsic curvature of $\sigma_0$ as a hypersurface in $S$
is defined by
\begin{equation}
  \bar{k}_{ab} = \frac12{}^{(3)}\bar{\pounds}_{\bar{n}}h_{ab},
  \label{def:bark_ab}
\end{equation}
where ${}^{(3)}\bar{\pounds}_{\bar{n}}$ is a Lie derivative
associated with $S$.

We now introduce the definition of a DTTS. 
From the above example of the Schwarzschild spacetime,
we consider the following definition to be
the most appropriate definition
of a DTTS:

%===========<DEFINITION>============%
%
\begin{definition}
  Suppose $\Sigma$ to be a smooth spacelike hypersurface of
  a spacetime $\mathcal{M}$. A closed orientable two-dimensional
  surface $\sigma_0$ in $\Sigma$ is a dynamically transversely trapping
  surface (DTTS) if and only if there exists a timelike hypersurface $S$
  in $\mathcal{M}$ that intersects $\Sigma$ precisely at $\sigma_0$ 
  and satisfies the following three conditions at arbitrary points on
  $\sigma_0$:
\begin{eqnarray}
  \bar{k} = 0, &\quad&
  \textrm{(the momentarily non-expanding condition);}
  \label{momentarily-non-expanding-condition}
  \\
  \mathrm{max}\left(\bar{K}_{ab}k^ak^b\right) = 0, &\quad&
  \textrm{(the marginally transversely trapping condition);}
  \label{marginally-transversely-trapping-condition}
  \\
  {}^{(3)}\bar{\pounds}_{\bar{n}} \bar{k}\le 0, &\quad&
  \textrm{(the accelerated contraction condition),} 
  \label{accelerated-contraction-condition}
\end{eqnarray}
where $k^a$ are arbitrary future-directed null vectors tangent to $S$ and
the quantity $\pounds_{\bar{n}}\bar{k}$ is evaluated
with a time coordinate in $S$ whose lapse function is constant
on $\sigma_0$.
\label{definition-1}
\end{definition}
%
%=================================%

Due to the inequality in the
condition in Eq.~\eqref{accelerated-contraction-condition}, 
if there is a DTTS in $\Sigma$, there are infinitely many DTTSs in $\Sigma$
in general. Using the concept of DTTSs, we introduce a
{\it dynamically transversely trapping region}
and a {\it marginally DTTS} by the following definitions:

%===========<DEFINITION>============%
%
\begin{definition}
  Consider a collection of DTTSs such that any two of these
  can be transformed to each other
  by continuous deformation keeping to DTTSs.
  The region in which these DTTSs exist is said to be a 
  dynamically transversely trapping region (or, more generally, 
  one of dynamically transversely trapping regions). If the
  outer boundary of a dynamically transversely trapping region
  satisfies
\begin{equation}
  {}^{(3)}\bar{\pounds}_{\bar{n}} \bar{k}= 0,
  \label{Equation-for-marginally-DTTS}
\end{equation}
  it is said to be a marginally DTTS.
\label{definition-2}
\end{definition}
%
%=================================%

Readers may wonder why 
we explicitly require Eq.~\eqref{Equation-for-marginally-DTTS}
for a marginally DTTS,
since Eq.~\eqref{Equation-for-marginally-DTTS} is naturally
expected to be satisfied by the fact that
a marginally DTTS is a boundary of a dynamically transversely
trapping region. This is 
because at least for now,
we cannot deny the possibility that the outer boundary is
given by an envelope of infinitely many DTTSs, and thus
it is not a DTTS.
Of course, it is quite uncertain whether such a case happens, 
and clarifying this point is interesting 
although it is beyond the scope of this paper.  
Note also that although Eq.~\eqref{Equation-for-marginally-DTTS} gives
an equation for marginally DTTSs, not all of the solutions
are marginally DTTSs, because some of them may be inner boundaries
of dynamically transversely trapping regions, or may be immersed
in those regions.

\subsection{Description of the three conditions}
\label{section2-3}

Among the three conditions in
Eqs.~\eqref{momentarily-non-expanding-condition}--\eqref{accelerated-contraction-condition},
the first two, the momentarily non-expanding condition of Eq.~\eqref{momentarily-non-expanding-condition}
and the marginally transversely trapping condition of Eq.~\eqref{marginally-transversely-trapping-condition}, determine the
behavior of $S$ in the neighborhood of $\sigma_0$.
The third condition, Eq.~\eqref{accelerated-contraction-condition}, is
to judge whether $\sigma_0$ exists in a strong gravity region. 
We discuss the meanings of the three
conditions in more detail, one by one.

\subsubsection{The momentarily non-expanding condition}
\label{section2-3-1}

There are infinitely many timelike hypersurfaces $S$ that intersect $\Sigma$
at $\sigma_0$, and the value of ${}^{(3)}\bar{\pounds}_{\bar{n}}\bar{k}$
strongly depends on the choice of $S$. 
For this reason, one must specify how to choose $S$, and 
the momentarily non-expanding condition is introduced
in order to fix the behavior of 
$S$ up to the first order in time.
The first-order behavior of $S$ is
specified by the timelike tangent vectors $\bar{n}^a$ of $S$
at points on $\sigma_0$, and once $\bar{n}^a$ is chosen,
the value of $\bar{k}$ is also determined because $\bar{k}_{ab}$ is
a quantity that represents the first-order behavior of $h_{ab}$ in time
as understood from Eq.~\eqref{def:bark_ab}.
Conversely, requiring $\bar{k}=0$ (basically) determines the choice
of $\bar{n}^a$, and thus the first-order behavior of $S$. 
To be specific, the tangent vector $\bar{n}^a$ can be given
in terms of $n^a$ and $\hat{r}^a$
through the transformation,
\begin{equation}
  \bar{n}^a = \frac{1}{\sqrt{1-\beta^2}}
  \left(n^a -\beta \hat{r}^a\right),
\end{equation}
where $\beta=\beta(x^i)$ is a function of the coordinates $x^i$ on $\sigma_0$.
The condition $\bar{k}=0$ is then becomes 
an equation for $\beta(x^i)$, and an appropriate $\bar{n}^a$
is obtained by solving this equation. 
Note that there are exceptional cases where $S$ is not uniquely specified
by this procedure. For example, if we adopt
time-symmetric initial data as $\Sigma$ and 
a minimal surface in $\Sigma$ as $\sigma_0$,
the value of $\bar{k}$ vanishes for arbitrary $\beta(x^i)$. 
In such cases, all surfaces $S$ with $\bar{k}=0$
must be taken into account.

The geometrical meaning of the momentarily non-expanding condition
is as follows. 
Let us span Gaussian normal coordinates $(t,x^i)$ in $S$  
from $\sigma_0$, and denote the slice $t=\mathrm{const.}$ as
$\sigma_t$. Here, $t$ is the time coordinate and $x^i$ are the
coordinates to specify positions of points on each of $\sigma_t$.
The momentarily non-expanding condition $\bar{k}=0$
means that each area element of $\sigma_t$ 
is unchanged up to the first order in time.
Therefore, the area of $\sigma_0$ is
locally extremal in a sequence of $\sigma_t$,
and if the accelerated contraction condition of Eq.~\eqref{accelerated-contraction-condition} is satisfied simultaneously, 
it is locally maximal.

\subsubsection{The marginally transversely trapping condition}
\label{section2-3-2}

The marginally transversely trapping condition
is introduced to determine the second-order behavior of $S$
in time using propagation of photons.

Consider a photon that is emitted from a point on $\sigma_0$
tangentially to $S$.
The quantity $\bar{K}_{ab}k^ak^b$ in Eq.~\eqref{marginally-transversely-trapping-condition}
represents whether 
the photon propagates in the inward direction of $S$ or not.
In order to see this, let us consider a ``virtual photon'' confined 
in the hypersurface $S$ whose equation is given by $k^a\bar{D}_{a}k^b=0$.
Rewriting with the four-dimensional quantities, we have
\begin{equation}
a^c=-(\bar{K}_{ab}k^ak^b)\hat{r}^c,
\end{equation}
where $a^c = k^d\nabla_dk^c$ is the four-acceleration. From this equation,
we understand that a real photon whose equation is given by
$k^a\nabla_{a}k^b=0$ propagates into the inward (resp. outward) region of $S$
if $\bar{K}_{ab}k^ak^b$ is negative (resp. positive).
If $\bar{K}_{ab}k^ak^b=0$, the photon travels on $S$
up to second order in time.
The condition 
$\bar{K}_{ab}k^ak^b=0$ is required
for a photon surface \cite{Claudel:2000},
while the condition $\bar{K}_{ab}k^ak^b\le 0$ is required for
a static/stationary TTS \cite{Yoshino:2017}.

The meaning of the marginally transversely trapping condition
in Eq.~\eqref{marginally-transversely-trapping-condition}
is as follows:
$S$ must be chosen so that 
all photons emitted from points on $\sigma_0$
in arbitrary tangential directions to $S$ must propagate
into the inside region of $S$ or precisely on $S$,
and furthermore, at least one photon must propagate 
on $S$ (up to the second order in time).
In other words, if we consider 
a collection of all such photons, they distribute
in a region with a small thickness in general 
after they are emitted. 
Then, $S$ is adopted as the outer boundary of such
a region.

\subsubsection{The accelerated contraction condition}
\label{section2-3-3}

Once the surface $S$ is specified by the above two conditions,
it is possible to calculate ${}^{(3)}\bar{\pounds}_{\bar{n}}\bar{k}$.
This is a quantity determined by the second-order behavior of $S$
in time. Here, we have to remark that this is a coordinate-dependent
quantity. The trace of the Ricci equation of $\sigma_0$
as a hypersurface in $S$ is 
\begin{equation}
  {}^{(3)}\bar{\pounds}_{\bar{n}}\bar{k}
  =-{}^{(3)}\bar{R}_{ab}\bar{n}^a\bar{n}^b
  -\bar{k}_{ab}\bar{k}^{ab}+\frac{1}{\alpha}\mathcal{D}^2\alpha,
  \label{trace-Ricci-sigma0-in-S}
\end{equation}
where ${}^{(3)}\bar{R}_{ab}$ is the Ricci tensor calculated
by the metric $\bar{p}_{ab}$ induced on $S$,
$\mathcal{D}^2:=\mathcal{D}_a\mathcal{D}^a$ 
is the Laplace operator on $\sigma_0$, and 
$\alpha$ is the lapse function of the time coordinate. Hence,
${}^{(3)}\bar{\pounds}_{\bar{n}}\bar{k}$ obviously depends on $\alpha$,
and thus we have to specify how to choose $\alpha$.
As remarked at the end of Definition~\ref{definition-1},
we require $\alpha=\mathrm{const.}$ on $\sigma_0$.
Then, the last term of Eq.~\eqref{trace-Ricci-sigma0-in-S}
vanishes and ${}^{(3)}\bar{\pounds}_{\bar{n}}\bar{k}$ is
given only in terms of geometrical quantities.

Note that, due to the constancy of $\alpha$, the concept
of a DTTS becomes different from that of
a static/stationary TTS. In the case of a static/stationary
TTS, we consider a static/stationary surface $S$
which is generated by the Lie drag of $\sigma_0$ along the integral lines
of the timelike Killing vector field $\xi^a$
in a static/stationary spacetime 
$\mathcal{M}$ \cite{Yoshino:2017}. Then, the
momentarily non-expanding 
condition of Eq.~\eqref{momentarily-non-expanding-condition} 
is trivially satisfied. If we choose $\sigma_0$ that
satisfies the marginally transversely
trapping condition of Eq.~\eqref{marginally-transversely-trapping-condition}, 
$S$ becomes a static/stationary TTS. However,
in this situation, although $S$ satisfies
${}^{(3)}\bar{\pounds}_{\bar{n}}\bar{k}=0$ for the time coordinate
associated with the timelike Killing vector field
(i.e. for the choice $\alpha = \sqrt{-\xi_a\xi^a}$),
${}^{(3)}\bar{\pounds}_{\bar{n}}\bar{k}$ violates 
the accelerated contraction condition of Eq.~\eqref{accelerated-contraction-condition}
in general for the choice $\alpha=\mathrm{const.}$,
except for spherically symmetric spacetimes.
Hence, there exist cases that a static/stationary TTS is not
a DTTS at the same time. Furthermore, as we will see
in Sect.~\ref{section5}, there are also converse cases that
a DTTS in a static spacetime is not a static TTS.
For this reason, an inclusion relationship cannot be found
between DTTSs and static/stationary TTSs.

\subsection{Comparison with trapped surfaces}
\label{section2-4}

A DTTS (Definition~\ref{definition-1})
and a marginally DTTS (Definition~\ref{definition-2})
are defined so that they have similarity to
a trapped surface and a marginally trapped surface, respectively.

A trapped surface $\sigma_{\rm TS}$
is a two-dimensional closed orientable surface
in a spacelike hypersurface $\Sigma$
such that both the outgoing and ingoing null geodesic congruences
have negative expansion,
i.e. $\theta_+< 0$ and $\theta_-< 0$, respectively.
A collection of trapped surfaces forms a trapped region,
and the outer boundary of the trapped region is defined as 
a marginally trapped surface. On the marginally trapped surface,  
$\theta_-\le 0$ and $\theta_+=0$ hold.
A marginally trapped surface may have multiple connected components
(among them, the outermost components are defined as the apparent horizons),
and the location of each component is uniquely determined. 
Similarly, a DTTS is defined in a spacelike hypersurface
$\Sigma$ and a collection of DTTSs forms a dynamically
transversely trapping region. A marginally DTTS
is defined in a similar manner as a marginally trapped surface,
and the location of each marginally DTTS 
is uniquely determined.

A (marginally) trapped surface and a (marginally) DTTS
have similar features of gauge invariance and gauge dependence.
On one hand, a (marginally) trapped surface $\sigma_{\rm TS}$
(seen as a two-dimensional spacelike surface in $\mathcal{M}$)
is a gauge-invariant concept
in the sense that values of the two kinds of
expansion, $\theta_+$ and $\theta_-$, do not depend
on the choice of coordinates to calculate them.
Suppose $\sigma_{\rm TS}$ is obtained as a (marginally) trapped surface
in some spacelike hypersurface $\Sigma$. Then, if (marginally) trapped
surfaces are surveyed on 
another spacelike hypersurface $\Sigma^\prime$ which intersects
with $\Sigma$ precisely at $\sigma_{\rm TS}$,
we would obtain $\sigma_{\rm TS}$ 
as a (marginally) trapped surface as well.
Similarly, if a surface $\sigma_0$ is an intersection
of two spacelike hypersurfaces $\Sigma$ and $\Sigma^\prime$,
we would obtain the same conclusion 
concerning whether $\sigma_0$ is a (marginally) DTTS or not
on both of the hypersurfaces, 
because the same timelike hypersurface $S$ should be constructed
by virtue of the momentarily non-expanding 
condition of Eq.~\eqref{momentarily-non-expanding-condition}
and the marginally transversely trapping
condition of Eq.~\eqref{marginally-transversely-trapping-condition}.
On the other hand, a marginally trapped surface has gauge-dependent feature
in the sense that if we consider time evolution of a marginally
trapped surface, its three-dimensional worldsheet,    
which is called the trapping horizon \cite{Hayward:1993}, depends on the
choice of the time coordinate. Similarly,
the worldsheet of a marginally DTTS
should be dependent on the selected time coordinate
as well.

Further similarity between a (marginally) trapped surface
and a (marginally) DTTS can be found for their area, i.e.
the Penrose inequality of Eq.~\eqref{Penrose-inequality}
and the Penrose-like inequality of Eq.~\eqref{Penroselike-inequality}.
This issue will be discussed in Sect.~\ref{section6}.

%
%======================================%
%<<<<<<<<<<<< SECTION III  >>>>>>>>>>>>>>%
%======================================%
%
\section{Configurations and useful formulas}
\label{section3}

In this paper, we solve for marginally DTTSs and discuss their properties
in fairly restricted situations. Namely, we consider the situation
where the timelike hypersurface $S$ that satisfies the
momentarily non-expanding condition in Eq.~\eqref{momentarily-non-expanding-condition}
orthogonally intersects with the spacelike hypersurface $\Sigma$.
In other words, we restrict our attention to the case
where
\begin{equation}
\hat{r}^a = \bar{r}^a \qquad \textrm{and} \qquad n^a=\bar{n}^a
\end{equation}
are satisfied. 
To be specific, we consider time-symmetric initial data and
momentarily stationary axisymmetric initial data.
First, we derive useful formulas that are applicable to these
setups in Sect.~\ref{section3-1}, and then 
the two kinds of initial data are described in Sect.~\ref{section3-2}.
Preparing the method of solving for (marginally) DTTSs
in the case that $S$ is not orthogonal to $\Sigma$
is definitely necessary, and we plan to study this issue
in a forthcoming paper.

\subsection{Useful formulas}
\label{section3-1}

Let us span the coordinates $(t, r, x^i)$ such that
the timelike hypersurface $S$ is given by $r=0$
in the neighborhood of $\sigma_0$. Let $\alpha$ denote
the time lapse function, and for simplicity,
we assume the shift vector to be zero, $\beta^a:=t^a-\alpha n^a=0$,
where $t^a$ is the basis of the coordinate $t$. The metric is given by
\begin{equation}
  ds^2 = -\alpha^2dt^2 + \varphi^2 dr^2 + h_{ij}dx^idx^j + 2\gamma_{ri}drdx^i,
  \label{metric-neighborhood-of-sigma0}
\end{equation}
where $\varphi$ is the lapse function of the radial coordinate $r$.

\subsubsection{Formulas related to the marginally transversely trapping condition}

First, we derive useful formulas to study the
marginally transversely trapping condition
of Eq.~\eqref{marginally-transversely-trapping-condition}.
Rewriting Eq.~\eqref{def:kab} for $k_{ab}$ in terms of four-dimensional
quantities, we have
\begin{equation}
k_{ab} = \frac12\pounds_{\hat{r}}h_{ab} + v_an_b+v_bn_a,
\end{equation}
with
\begin{equation}
  v_a = \frac12h_{ab}\pounds_{n}\hat{r}^b.
  \label{va_expression1}
\end{equation}
By substituting $\bar{p}_{ab}$ given by Eq.~\eqref{def:barpab}
into the formula for $\bar{K}_{ab}$, Eq.~\eqref{def:barKab}, we obtain
\begin{equation}
  \bar{K}_{ab} = -n_an_b\frac{{}^{(3)}\pounds_{\hat{r}}\alpha}{\alpha} +
  \frac12\pounds_{\hat{r}}h_{ab}
\end{equation}
after some algebra. Therefore, we obtain the following decomposition
of $\bar{K}_{ab}$ into time and space directions:
\begin{equation}
  \bar{K}_{ab} = -n_an_b\frac{{}^{(3)}\pounds_{\hat{r}}\alpha}{\alpha}
  +k_{ab}-v_an_b-v_bn_a.
  \label{barKab-decomposition}
\end{equation}
There are several other expressions for $v_a$, for example, 
\begin{equation}
  v_a=h_{ab}n^c\nabla_c\hat{r}^b,
  \label{va_expression2}
\end{equation}
which is derived by expressing the right-hand side with
$\bar{K}_{d}^{~b}=\bar{p}_{d}^{~c}\nabla_c\hat{r}^b$ and
substituting Eq.~\eqref{barKab-decomposition}.
From Eqs.~\eqref{va_expression1} and \eqref{va_expression2}, we have
\begin{equation}
  v_a = -h_{ab}\hat{r}^c\nabla_cn^b
  =-h_{ab}\hat{r}^dK_{d}^{~b},
  \label{va_expression3}
\end{equation}
where $K_{d}^{~b}=q_{d}^{~c}\nabla_cn^b$ is used in the second equality.
The null tangent vector $k^a$ of $S$
that appears in the marginally transversely trapping
condition of Eq.~\eqref{marginally-transversely-trapping-condition}
is expressed as 
\begin{equation}
k^a = n^a+s^a,
\end{equation}
using the timelike unit normal $n^a$ to $\Sigma$
and unit tangent vectors $s^a$ to $\sigma_0$.
Then, the marginally transversely trapping
condition in Eq.~\eqref{marginally-transversely-trapping-condition}
is rewritten as
\begin{equation}
  \mathrm{max}(k_{ab}s^as^b+2v_bs^b)=\frac{{}^{(3)}\pounds_{\hat{r}}\alpha}{\alpha}.
  \label{marginally-transversely-trapping-condition-rewritten}
\end{equation}

\subsubsection{Formulas related to the accelerated contraction condition}

Next, we obtain a useful formula for
calculating ${}^{(3)}\bar{\pounds}_{\bar{n}}\bar{k}$
in the accelerated contraction condition
of Eq.~\eqref{accelerated-contraction-condition}.
We recall the trace of the Ricci equation, Eq.~\eqref{trace-Ricci-sigma0-in-S},
on $\sigma_0$ as a hypersurface in $S$,
and rewrite with the double trace of the Gauss
equation on $\sigma_0$ in $S$,
\begin{equation}
  {}^{(2)}R = {}^{(3)}\bar{R} + 2{}^{(3)}\bar{R}_{ab}n^an^b
  -\bar{k}^2+\bar{k}_{ab}\bar{k}^{ab},
\end{equation}
where ${}^{(2)}R$ is the Ricci scalar of $\sigma_0$, and the
double trace of the Gauss equation on $S$ in the spacetime $\mathcal{M}$,
\begin{equation}
{}^{(3)}\bar{R} = -2G_{ab}\hat{r}^a\hat{r}^b+\bar{K}^2-\bar{K}_{ab}\bar{K}^{ab},
\end{equation}
where $G_{ab}:=R_{ab}-(1/2)g_{ab}R$ is the Einstein tensor.
The result is
\begin{equation}
  {}^{(3)}\bar{\pounds}_{n}\bar{k} = -\frac12{}^{(2)}R
  -8\pi GP_r
  +\frac12\left(\bar{K}^2-\bar{K}_{ab}\bar{K}^{ab}-\bar{k}^2-\bar{k}_{ab}\bar{k}^{ab}\right)
  +\frac{1}{\alpha}\mathcal{D}^2\alpha,
  \label{Lie-derivative-of-bark-rewritten}
\end{equation}
where the Einstein field equations $G_{ab}=8\pi GT_{ab}$ are assumed
with the energy-momentum tensor $T_{ab}$,
and the radial pressure is introduced by
$P_{r} := T_{ab}\hat{r}^a\hat{r}^b$.
%  \label{def:radial-pressure}
For a later convenience, we also define the energy density
by $\rho:=T_{ab}n^an^b$.

We evaluate this equation on a surface $\sigma_0$
which is to be examined as to whether it is a DTTS. The
last term vanishes because $\alpha=\mathrm{const.}$
is required from Definition~\ref{definition-1},
and $\bar{k}=0$ from the momentarily non-expanding condition
of Eq.~\eqref{momentarily-non-expanding-condition}.
Substituting the decomposed form of $\bar{K}_{ab}$,
Eq.~\eqref{barKab-decomposition},
into Eq.~\eqref{Lie-derivative-of-bark-rewritten},
we have
\begin{equation}
{}^{(3)}\bar{\pounds}_{n}\bar{k} = -\frac12{}^{(2)}R
-8\pi GP_r
+k\frac{{}^{(3)}\pounds_{\hat{r}}\alpha}{\alpha}
+\frac12\left(k^2-k_{ab}k^{ab}-\bar{k}_{ab}\bar{k}^{ab}\right)
+v_av^a.
\label{Lie-derivative-bark-final-form}
\end{equation}
The quantity $\bar{k}_{ab}$ in this formula can be
calculated from $K_{ab}$, because similarly to Eq.~\eqref{barKab-decomposition},
$K_{ab}$ is decomposed as
\begin{equation}
  K_{ab} = \hat{r}_a\hat{r}_b\frac{{}^{(3)}\bar{\pounds}_{n}\varphi}{\varphi}
  +\bar{k}_{ab}-v_a\hat{r}_b-v_b\hat{r}_a,
  \label{Kab-decomposition}
\end{equation}
where $v_a$ is given in Eq.~\eqref{va_expression1}, and thus
\begin{equation}
\bar{k}_{ab}=h_{a}^{~c}h_{b}^{~d}K_{cd}
\label{barkab-Kcd}
\end{equation}
holds.

\subsection{Configurations}
\label{section3-2}

We now describe two specific configurations to be studied
in this paper, the time-symmetric
initial data and the momentarily stationary axisymmetric initial data. 

\subsubsection{Time-symmetric initial data}

Initial data are said to be time symmetric or
momentarily static if it has vanishing extrinsic curvature,
\begin{equation}
  K_{ab}=0.
\label{Extrinsic-curvature:time-symmetric}
\end{equation}
From Eq.~\eqref{barkab-Kcd} we have $\bar{k}_{ab}=0$, and therefore
the timelike hypersurface $S$ certainly satisfies the
momentarily non-expanding condition
of Eq.~\eqref{momentarily-non-expanding-condition},
$\bar{k}=0$. From Eq.~\eqref{va_expression3}, we have $v_a=0$.
To find the expression for the marginally transversely trapping
condition of Eq.~\eqref{marginally-transversely-trapping-condition} in this setup,
it is convenient to choose the orthonormal basis
$\mathbf{e}_1$ and $\mathbf{e}_2$ on $\sigma_0$ which diagonalizes 
$k_{ab}$ as
\begin{equation}
k_{ab} = k_{\rm 1}(\mathbf{e}_1)_a(\mathbf{e}_1)_b
+k_{\rm 2}(\mathbf{e}_2)_a(\mathbf{e}_2)_b,
\label{kab-dagonalized}
\end{equation}
and define
\begin{subequations}
\begin{eqnarray}
  k_{\rm L} &=& \mathrm{max}(k_1,k_2), \label{def:k_L}\\
  k_{\rm S} &=& \mathrm{min}(k_1,k_2). \label{def:k_S}
\end{eqnarray}
\end{subequations}
Then, from Eq.~\eqref{marginally-transversely-trapping-condition-rewritten},
the marginally transversely trapping condition becomes
\begin{equation}
  k_{\rm L} = \frac{{}^{(3)}\pounds_{\hat{r}}\alpha}{\alpha}.
  \label{marginally-transversely-trapping-condition-time-symmetric}
\end{equation}
Using this formula with $\bar{k}_{ab}=0$, we obtain
\begin{equation}
2{}^{(3)}\bar{\pounds}_{n}\bar{k} = -{}^{(2)}R
-16\pi GP_r
+2kk_{\rm L}
+k^2-k_{ab}k^{ab}.
\label{Lie-derivative-bark-time-symmetric}
\end{equation}
This equation is used in solving for the marginally
DTTS in Sects.~\ref{section4} and \ref{section5},
and in studying the Penrose-like inequality in Sect.~\ref{section6}.

\subsubsection{Momentarily stationary axisymmetric initial data}

The initial data $\Sigma$ is said to be momentarily stationary
if, for an appropriately chosen time coordinate $\tilde{t}$
with the basis $\tilde{t}^a=\tilde{\alpha}n^a+\tilde{\beta}^a$,
the Lie derivative of the metric $q_{ab}$ induced on $\Sigma$
with respect to $\tilde{t}^a$ vanishes,
\begin{equation}
  \pounds_{\tilde{t}}q_{ab}
  =2\tilde{\alpha}K_{ab} + D_a\tilde{\beta}_b + D_b\tilde{\beta}_a =0.
  \label{momentarily-stationary-condition}
\end{equation}
Note that $\tilde{t}^a$, $\tilde{\alpha}$, and $\tilde{\beta}^a$
are different from $t^a$, $\alpha$, and $\beta^a$ introduced
just before Eq.~\eqref{metric-neighborhood-of-sigma0}.
At the same time, $\Sigma$ is assumed to be axisymmetric
with the Killing vector $\phi^a$, and the azimuthal angular coordinate $\phi$
is introduced by $\phi^a=(\partial_\phi)^a$. We further
assume that the shift vector $\tilde{\beta}^a$ is given in the form
\begin{equation}
  \tilde{\beta}^a = -\tilde{\omega} \phi^a,
  \label{conditon-for-beta}
\end{equation}
where $\tilde{\omega}$ does not depend on $\phi$ and satisfies
${}^{(3)}\pounds_{\phi}\tilde{\omega}=0$.
Then, from Eqs.~\eqref{momentarily-stationary-condition}
and \eqref{conditon-for-beta}, the extrinsic curvature
is 
\begin{equation}
  K_{ab} = \frac{1}{2\tilde{\alpha}}\left[(D_a\tilde{\omega})\phi_b+(D_b\tilde{\omega})\phi_a\right],
\label{Extrinsic-curvature:momentarily-stationary-axisymmetric}
\end{equation}
where the Killing equation $D_{(a}\phi_{b)}=0$ is used.
We define $\Sigma$ to be momentarily stationary axisymmetric initial
data if the extrinsic curvature is given in the form of Eq.~\eqref{Extrinsic-curvature:momentarily-stationary-axisymmetric}.
Note that $\tilde{\alpha}$ and $\tilde{\omega}$ are not uniquely determined.
If we consider the transformation
\begin{equation}
\tilde{\omega}^\prime = g(\tilde{\omega}), \qquad
\tilde{\alpha}^\prime = \frac{dg}{d\tilde{\omega}}\ \alpha,
\end{equation} 
where $g(\tilde{\omega})$ is an arbitrary monotonically increasing function,
the same extrinsic curvature is given in the same form as  
Eq.~\eqref{Extrinsic-curvature:momentarily-stationary-axisymmetric}
but with $\tilde{\alpha}$ and $\tilde{\omega}$ replaced by
$\tilde{\alpha}^\prime$ and $\tilde{\omega}^\prime$, respectively.
From this construction, 
the spacetime $\mathcal{M}$ possesses the symmetry under the simultaneous
transformations $\tilde{t}\to -\tilde{t}$ and $\phi\to -\phi$.
The structure of the initial data 
$\Sigma$ is invariant under the transformation $\phi\to -\phi$.

An example of momentarily stationary axisymmetric initial data
is a $t=\mathrm{const.}$ hypersurface of a Kerr spacetime
in the Boyer-Lindquist coordinates. Various other configurations
can be considered. If matter distributes in an axially symmetric manner
with an axially symmetric velocity field 
directed in the $\phi$ direction,
the initial data become momentarily stationary and axisymmetric.
See, e.g., Ref.~\cite{Nakao:2014} for a numerically constructed example.

We adopt an axisymmetric 
surface as $\sigma_0$. Then, 
$\phi^a$ is a tangent vector to $\sigma_0$
and satisfies $\phi_a=h_{ab}\phi^b$.
Equation~\eqref{barkab-Kcd} implies that
\begin{equation}
  \bar{k}_{ab} = \frac{1}{2\tilde{\alpha}}\left[(\mathcal{D}_a\tilde{\omega})\phi_b+(\mathcal{D}_b\tilde{\omega})\phi_a\right],
\label{barkab:momentarily-stationary-axisymmetric}
\end{equation}
and thus we have $\bar{k}=0$ by virtue of the axisymmetry of $\sigma_0$.
Therefore, the momentarily non-expanding
condition in Eq.~\eqref{momentarily-non-expanding-condition} is satisfied.

Let us examine the metric in the form of Eq.~\eqref{metric-neighborhood-of-sigma0}
in the coordinates $(t, r, x^i)$.
We set $x^1=\theta$ and $x^2=\phi$, where
$\theta$ and $\phi$ are polar and azimuthal coordinates,
respectively.
From the symmetry under the transformation $\phi\to -\phi$,
it is possible to introduce $\theta$ so that $h_{\theta\phi}=0$ holds
on $\Sigma$. The surface $\sigma_0$ is supposed to be given by $r=0$,
and the angular coordinates can be spanned to satisfy 
$\gamma_{r\theta}=h_{r\phi}=0$ in the vicinity of $\sigma_0$
on $\Sigma$. Under this situation, the nonzero metric functions
are $\alpha$, $\varphi$, $h_{\theta\theta}$, and $h_{\phi\phi}$,
and from Eq.~\eqref{Extrinsic-curvature:momentarily-stationary-axisymmetric}, 
only the $r\phi$ and $\theta\phi$ components of the extrinsic
curvature $K_{ab}$ are nonzero. Therefore,
the functions $\alpha$, $\varphi$, $h_{\theta\theta}$, and $h_{\phi\phi}$
behave as even functions while 
$\gamma_{r\phi}$ and $h_{\theta\phi}$ behave as odd functions
with respect to $t$. It is convenient to introduce the
orthonormal basis on $\sigma_0$ by
\begin{equation}
  \mathbf{e}_1 = \sqrt{h_{\theta\theta}}\ \mathrm{d}\theta,
  \qquad
  \mathbf{e}_2 = \sqrt{h_{\phi\phi}}\ \mathrm{d}\phi,
\end{equation}
where the operator ``$\mathrm{d}$'' denotes the external derivative. 
Due to the symmetry under the transformation $\phi\to -\phi$,
$k_{ab}$ is given in the diagonalized
form of Eq.~\eqref{kab-dagonalized} with this orthonormal basis.

We now examine the marginally transversely trapping
condition of Eq.~\eqref{marginally-transversely-trapping-condition}.
Substituting Eq.~\eqref{Extrinsic-curvature:momentarily-stationary-axisymmetric}
into Eq.~\eqref{va_expression3}, we have
\begin{equation}
  v_a=-\frac{1}{2\tilde{\alpha}}({}^{(3)}\pounds_{\hat{r}}\tilde{\omega})\phi_a.
  \label{va:momentarily-stationary-axisymmetric}
\end{equation}
In terms of the orthonormal basis, $v_a$ is given by
$v_a=v_1(\mathbf{e}_1)_a+v_2(\mathbf{e}_2)_a$ with
\begin{equation}
v_1=0, \qquad v_2 = -\frac{1}{2\tilde{\alpha}}
  ({}^{(3)}\pounds_{\hat{r}}\tilde{\omega})\sqrt{\phi^a\phi_a}.
\end{equation}
Then, the marginally transversely trapping condition of Eq.~\eqref{marginally-transversely-trapping-condition-rewritten}
is rewritten as
\begin{equation}
\mathrm{max}\left(k_1,k_2+2|v_2|\right) = \frac{{}^{(3)}\pounds_{\hat{r}}\alpha}{\alpha}.
\label{marginally-transversely-trapping-condition-momentarily-stationary-axisymmetric-case}
\end{equation}
If we introduce $k_{\rm L}$ and $k_{\rm S}$ in the same manner as
Eqs.~\eqref{def:k_L} and \eqref{def:k_S}, the
inequality
\begin{equation}
  k_{\rm L}\le \frac{{}^{(3)}\pounds_{\hat{r}}\alpha}{\alpha}
  \label{inequality-Lr-alpha-momentarily-stationary-axisymmetric}
\end{equation}
is satisfied. From Eqs.~\eqref{Lie-derivative-bark-final-form},
\eqref{barkab:momentarily-stationary-axisymmetric}, and
\eqref{va:momentarily-stationary-axisymmetric}, 
the quantity ${}^{(3)}\bar{\pounds}_{n}\bar{k}$ that
appears in the accelerated contraction condition becomes
\begin{equation}
2{}^{(3)}\bar{\pounds}_{n}\bar{k} = -{}^{(2)}R
-16\pi GP_r
+2k\frac{{}^{(3)}\pounds_{\hat{r}}\alpha}{\alpha}
+k^2-k_{ab}k^{ab}
+\frac{(\phi^a\phi_a)}{2\tilde{\alpha}^2}
\left[({}^{(3)}\pounds_{\hat{r}}\tilde{\omega})^2-(\mathcal{D}\tilde{\omega})^2\right],
\label{Lie-derivative-bark-momentarily-stationary-axisymmetric}
\end{equation}
where ${}^{(3)}\pounds_{\hat{r}}\alpha/\alpha$ must be evaluated with Eq.~\eqref{marginally-transversely-trapping-condition-momentarily-stationary-axisymmetric-case}. 
This equation is used in studying the Penrose-like inequality in Sect.~\ref{section6}.
Although explicitly 
constructing marginally DTTSs in momentarily stationary axisymmetric initial
data is an interesting problem, we postpone it as a future work.

%
%======================================%
%<<<<<<<<<<<< SECTION IV  >>>>>>>>>>>>>>%
%======================================%
%
\section{Explicit examples in Brill--Lindquist initial data}
\label{section4}

In this section, we explicitly construct marginally DTTSs
in the Brill--Lindquist initial data \cite{Brill:1963}.
In Sect.~\ref{section4-1}, we explain the Brill--Lindquist initial
data and our setups. The equation for solving a marginally DTTS
is explained in Sect.~\ref{section4-2}, and numerical results
are presented in Sect.~\ref{section4-3}. Comparison with
marginally trapped surfaces is made in Sect.~\ref{section4-4}.

\subsection{Setup}
\label{section4-1}

The Brill--Lindquist initial data are time-symmetric asymptotically flat 
initial data with vanishing extrinsic curvature, $K_{ab}=0$,
and with conformally flat geometry,
\begin{equation}
ds^2 = \varPsi^4(dx^2+dy^2+dz^2).
\end{equation}
The momentum constraint is trivially satisfied and
the Hamiltonian constraint is reduced to
\begin{equation}
  \bar{\nabla}^2 \varPsi =0,
\end{equation}
for a vacuum spacetime, 
where $\bar{\nabla}^2$ denotes the flat space Laplacian,
\begin{equation}
  \bar{\nabla}^2 = \partial_x^2 + \partial_y^2 + \partial_z^2.
  \label{flat-space-laplacian}
\end{equation}
In general, the Brill--Lindquist initial data represent
$N$ black holes momentarily at rest. Here, we focus our attention
on the two equal-mass black holes and
choose the following solution to Eq.~\eqref{flat-space-laplacian},
\begin{equation}
\varPsi = 1+\frac{GM}{4\tilde{r}_+}+\frac{GM}{4\tilde{r}_-},
\end{equation}
with
\begin{equation}
  \tilde{r}_{\pm} = \sqrt{x^2+y^2+(z\mp z_0)^2},
  \label{def:rpm}
\end{equation}
where $M$ corresponds to the ADM mass. 
In the case of $z_0=0$, these initial data represent
an Einstein-Rosen bridge of a Schwarzschild spacetime.
By contrast, in the case of $z_0>0$, 
two Einstein-Rosen bridges are present in these initial data 
and the points $(x,y,z)=(0,0,\pm z_0)$ correspond to
two asymptotically flat regions beyond the bridges.

\subsection{The equation for a marginally DTTS}
\label{section4-2}

There are two kinds of marginally DTTSs
in these initial data: a marginally DTTS that encloses both
black holes (hereafter, a common marginally DTTS),
and two marginally DTTSs each of which encloses one of the
two black holes.

In order to solve for a common marginally DTTS $\sigma_0$, 
we introduce the spherical-polar coordinates
$(\tilde{r}, \tilde{\theta}, \phi)$ in the ordinary manner
as
\begin{subequations}
\begin{eqnarray}
  x &=& \tilde{r}\sin\tilde{\theta}\cos\phi,
  \label{spherical-polar-x}
  \\
  y&=& \tilde{r}\sin\tilde{\theta}\sin\phi,
  \label{spherical-polar-y}
  \\
  z&=& \tilde{r}\cos\tilde{\theta},
  \label{spherical-polar-z}
\end{eqnarray}
\end{subequations}
and give the surface $\sigma_0$ as $\tilde{r}=h(\tilde{\theta})$.
In order to derive an equation for a common marginally DTTS,
we introduce new coordinates $(r,\theta)$
in the vicinity of $\sigma_0$, where
\begin{subequations}
\begin{eqnarray}
  \tilde{r} &=& r+h(\theta),
  \label{coordinate-transformation-tilder-r-theta}
  \\
  \tilde{\theta} &=& \theta-p(r,\theta).
  \label{coordinate-transformation-tildetheta-r-theta}
\end{eqnarray}
\end{subequations}
We require $\tilde{\theta}=\theta$ on $\sigma_0$ (that is, $p(0,\theta)=0$),
and in these coordinates, $\sigma_0$ is given by $r=0$, consistently
with Eq.~\eqref{metric-neighborhood-of-sigma0}.
Then, the induced metric on $\sigma_0$
becomes
\begin{equation}
  ds^2 = \varPsi^4\left[(h^2+h^{\prime 2})d\theta^2 + h^2\sin^2\theta d\phi^2\right],
  \label{induced-metric-sigma0-BL}
\end{equation}
and we introduce the orthonormal basis on $\sigma_0$ as
\begin{equation}
  \mathbf{e}_1 = \varPsi^2\sqrt{h^2+h^{\prime 2}}\ \mathrm{d}\theta,
  \qquad
  \mathbf{e}_2 = \varPsi^2 h\sin\theta\ \mathrm{d}\phi.
\end{equation}
With this basis, $k_{ab}$ is diagonalized
in the form of Eq.~\eqref{kab-dagonalized}
by axial symmetry of this system.
Requiring
${}^{(3)}\bar{\pounds}_{n}\bar{k} =0$
in Eq.~\eqref{Lie-derivative-bark-time-symmetric},
the equation for a marginally DTTS is given by
\begin{equation}
  -\frac12{}^{(2)}R -8\pi GP_{r} + k_1k_2+(k_1+k_2)\mathrm{max}(k_1,k_2)=0, 
  \label{Equation-for-a-marginally-DTTS-time-symmetric}
\end{equation}
where $P_{r}=0$
in the setup of the vacuum Brill--Lindquist initial data here.
We must express this equation as the equation for $h(\theta)$,
and this procedure is presented in Appendix~\ref{Appendix-A}.
Note that the cases $k_1\le k_2$ and $k_1\ge k_2$ must be
studied separately, and in both cases, Eq.~\eqref{Equation-for-a-marginally-DTTS-time-symmetric}
becomes second-order ordinary differential equations for $h(\theta)$.
These equations are solved using the fourth-order Runge--Kutta method
under the boundary condition $h^\prime=0$ 
at $\theta=0$ and $\pi/2$. 
In the case of common marginally DTTSs, 
the equation for the case $k_1\le k_2$
is solved, and as a result, a DTTS that satisfies  $k_1\le k_2$
is obtained consistently.

%===========<FIGURE1>============%
%
\begin{figure}[tb]
\centering
\includegraphics[width=0.5\textwidth,bb=0 0 260 252]{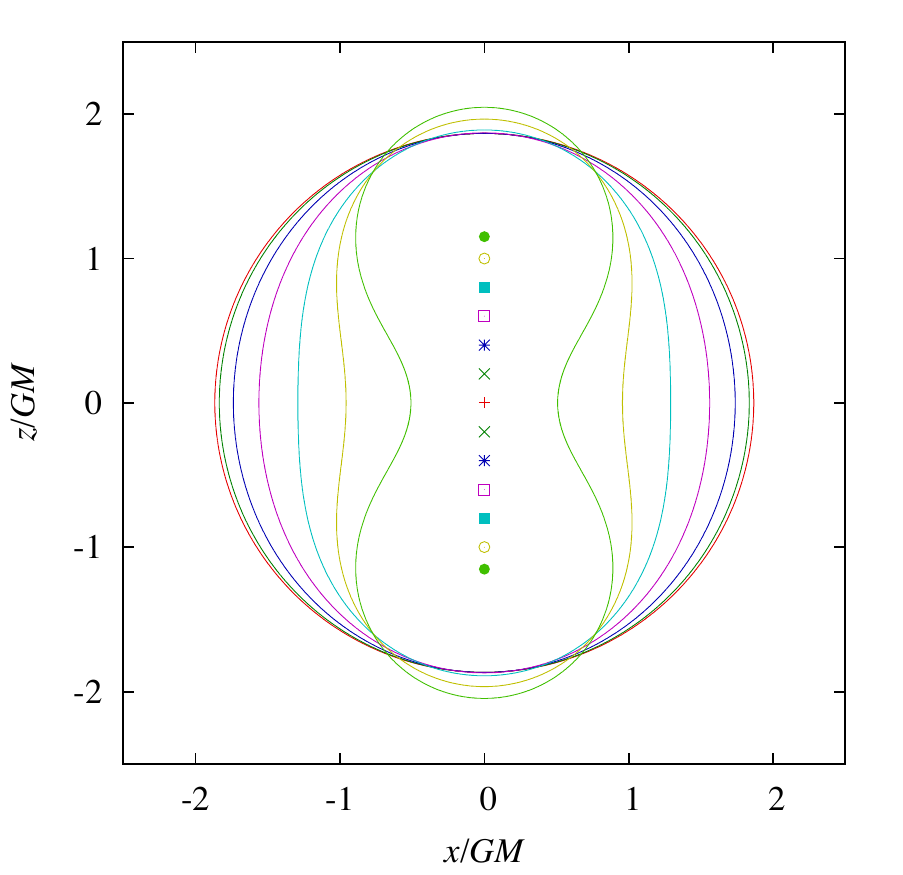}
\caption{
  Sections of common marginally DTTSs with the $(x,z)$-plane
  in the Brill--Lindquist initial data
  for $z_0/GM=0.0$, $0.2$, $0.4$, $0.8$, $1.0$, and $1.1506$.
  For $z_0/GM\ge 1.1507$, no common marginally DTTS is present.  
}
\label{BL-common-all}
\end{figure}
%
%=================================%

%===========<FIGURE1>============%
%
\begin{figure}[tb]
\centering
\includegraphics[width=0.96\textwidth,bb=0 0 454 169]{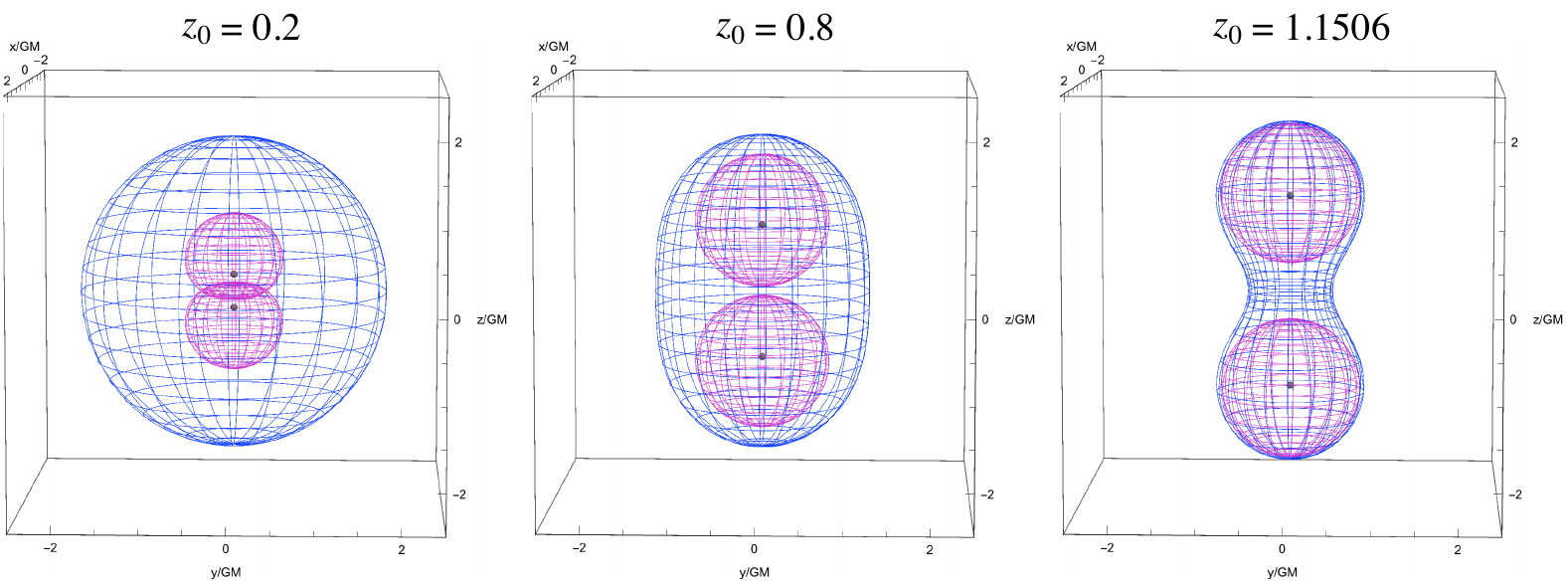}
\caption{
  3D plots of marginally DTTSs in the Brill--Lindquist initial data
  for $z_0/GM=0.2$ (left panel), $0.8$ (middle panel),
  and $1.1506$ (right panel). In each panel, three marginally DTTSs
  are presented: one is the common DTTS, and each of
  the other two DTTSs encloses only one black hole.   
}
\label{BL-common-each-TTSs}
\end{figure}
%
%=================================%

In solving for a marginally DTTS that encloses only one of the two black holes,
we introduce the spherical-polar coordinates by Eqs.~\eqref{spherical-polar-x},
\eqref{spherical-polar-y}, and 
\begin{equation}
  z\ =\ z_0+\tilde{r}\cos\tilde{\theta},
  \label{spherical-polar-z2}
\end{equation}
instead of Eq.~\eqref{spherical-polar-z}, so that 
the coordinate origin $\tilde{r}=0$ corresponds to $(x,y,z)=(0,0,z_0)$.
Parametrizing a marginally DTTS as $\tilde{r}=h(\tilde{\theta})$,
the new coordinates $(r,\theta)$ are introduced in the same manner
as Eqs.~\eqref{coordinate-transformation-tilder-r-theta} and
\eqref{coordinate-transformation-tildetheta-r-theta}. Then, 
the equations have the same form as the case of the common marginally DTTSs.
The boundary condition is $h^\prime=0$ at $\theta=0$ and $\pi$. 
Solving the equation for the case $k_1\ge k_2$,
a DTTS that satisfies $k_1\ge k_2$ is consistently obtained.

\subsection{Numerical results}
\label{section4-3}

We now show the numerical results. Figure~\ref{BL-common-all}
shows the sections of the common marginally DTTSs
with the $(x,z)$-plane 
in the Brill--Lindquist initial data for
$z_0/GM=0.0$, $0.2$, $0.4$, $0.8$, $1.0$, and $1.1506$.
For $z_0/GM=0.0$, the common marginally DTTS is spherically
symmetric with radius $\tilde{r}/GM=1+\sqrt{3}/2$. This corresponds
to the radius of the photon sphere of a Schwarzschild spacetime
in the isotropic coordinates. The common DTTS becomes distorted
as the value of $z_0/GM$ is increased, 
and it exists up to  $z_0/GM\approx 1.1506$. 
We could not find the common DTTS for $z_0/GM\ge 1.1507$
for the following reason. 
For the parameter region $1.007\lesssim z_0/GM\lesssim 1.1506$,
we obtain two solutions that correspond to the
outer and inner boundaries of the (common) dynamically transversely
trapping region that encloses both black holes.\footnote{We could
  not obtain a solution of the inner boundary for $0<z_0/GM\lesssim 1.006$
  because $h(\theta)$ becomes a multi-valued function
  for this parameter range. This is a technical problem and
  the inner boundary would also exist for this range of $z_0/GM$.}
Here, the outer boundary is the
marginally DTTS, and
the inner boundary is not depicted in Fig.~\ref{BL-common-all}.
Around $z_0/GM\approx 1.1506$, the outer 
and inner boundaries degenerate
and the common dynamically transversely
trapping region becomes infinitely thin, and it vanishes as 
$z_0/GM$ is further increased.

%===========<FIGURE1>============%
%
\begin{figure}[tb]
\centering
\includegraphics[width=0.9\textwidth,bb=0 0 416 144]{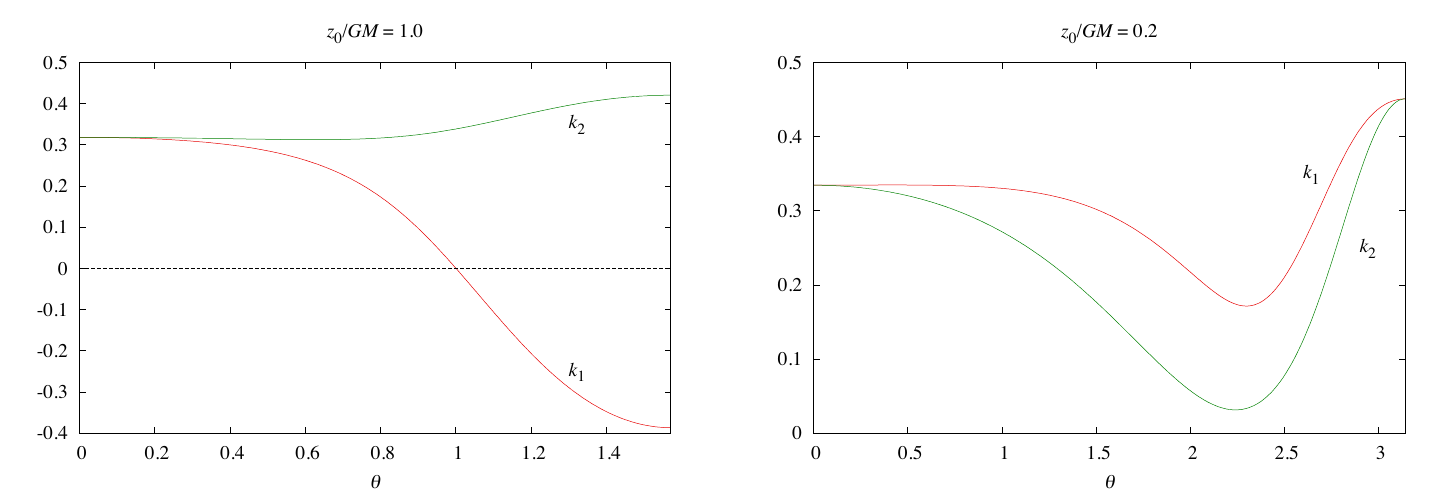}
\caption{
  Behavior of $k_1$ and $k_2$ on marginally DTTSs as functions of the polar
  angle $\theta$ in the Brill--Lindquist case. Left panel:
  $k_1$ and $k_2$ for the common marginally DTTS in the case $z_0/GM=1.0$.
  The relation $k_1\le k_2$ is kept on the surface.
  Right panel: $k_1$ and $k_2$ for the marginally DTTS that encloses
  only the upper black hole in the case $z_0/GM=0.2$.
  The relation $k_1\ge k_2$ is kept on the surface.
}
\label{k1_k2_BL}
\end{figure}
%
%=================================%

Figure~\ref{BL-common-each-TTSs} shows three-dimensional (3D) plots of the marginally DTTSs
for the cases $z_0/GM=0.2$ (left panel), $0.8$ (middle panel),
and $1.1506$ (right panel). In this figure,
two kinds of marginally DTTSs are plotted: one is
the common marginally DTTS and the other two are the
marginally DTTSs each of which surrounds only one of the two black holes.
For small $z_0/GM$, the two inner marginally DTTSs cross with each
other as shown in the left panel.
Therefore, there are cases where two dynamically
transversely trapping regions overlap.  
As $z_0/GM$ is increased, the two dynamically
transversely trapping regions become separate
as shown in the middle and right panels.
For $z_0/GM\ge 1.1507$, no common DTTS 
can be found, but two separate marginally DTTSs can always be found.

Figure~\ref{k1_k2_BL} shows the values of $k_1$ and $k_2$
as functions of the polar angle $\theta$
for a common DTTS in the initial data with $z_0/GM=1.0$
(left panel) and for a DTTS that surrounds only the upper
black hole (right panel) in the initial data with $z_0/GM=0.2$.
Because the equation for a marginally DTTS depends on the
sign of $k_1-k_2$, whether the behavior of $k_1$ and $k_2$
is consistent with the chosen equation 
must be
checked after the solutions are obtained.
For a common DTTS, the relation $k_1\le k_2$ is kept,
while for a DTTS that surrounds only the upper black hole,
the relation $k_1\ge k_2$ is kept.

\subsection{Comparison with marginally trapped surfaces}
\label{section4-4}

%===========<FIGURE1>============%
%
\begin{figure}[tb]
\centering
\includegraphics[width=0.5\textwidth,bb=0 0 270 252]{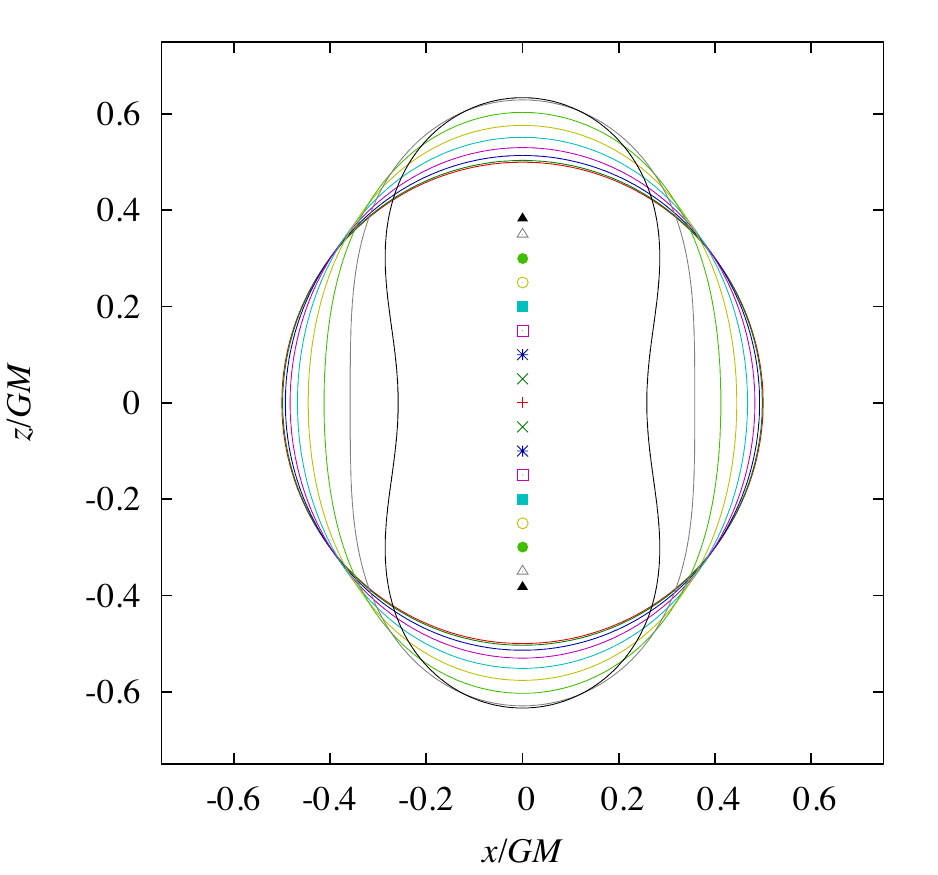}
\caption{
  Sections of common apparent horizons with the $(x,z)$-plane
  in the Brill--Lindquist initial data
  for $z_0/GM=0.0$, $0.05$, $0.10$, $0.15$, $0.20$, $0.25$,
  $0.30$, $0.35$, and $0.3830$.
  For $z_0/GM\ge 0.3831$, no common apparent horizon is present.
  Compare with Fig.~\ref{BL-common-all}.
}
\label{BL-common-AH-all}
\end{figure}
%
%=================================%

%===========<FIGURE1>============%
%
\begin{figure}[tb]
\centering
\includegraphics[width=0.96\textwidth,bb=0 0 454 167]{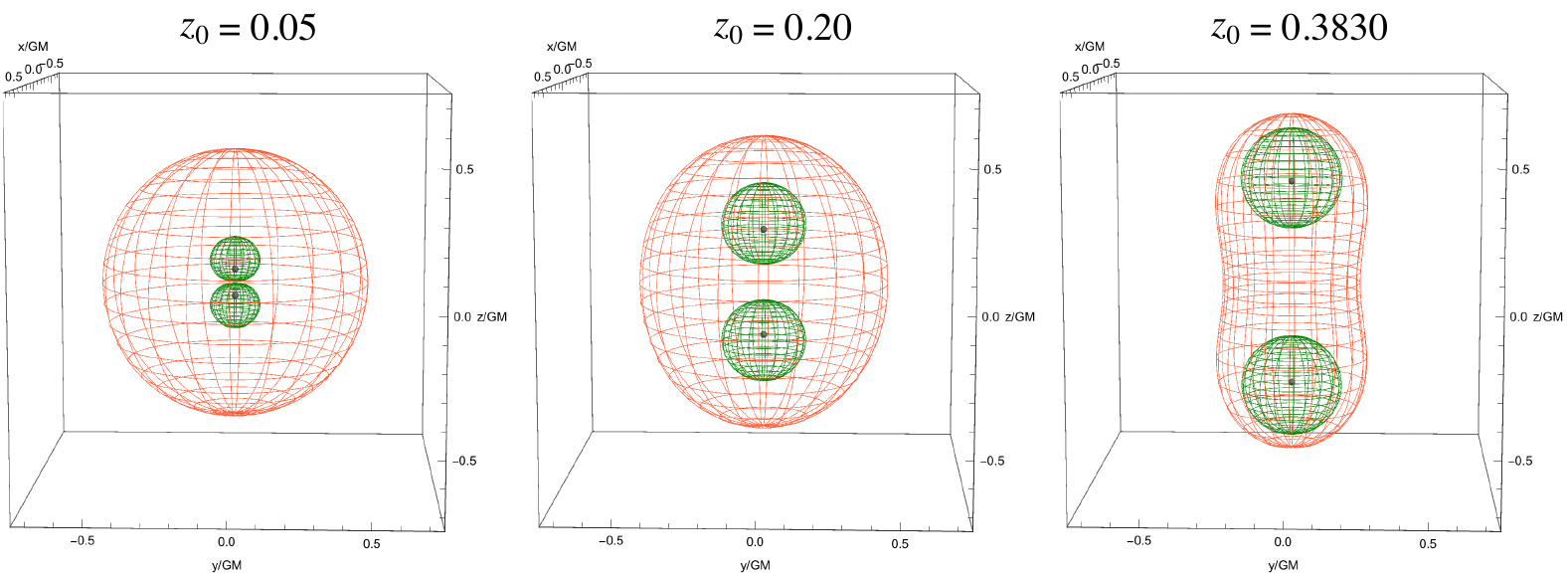}
\caption{
  3D plots of marginally trapped surfaces in the Brill--Lindquist initial data
  for $z_0/GM=0.05$ (left panel), $0.20$ (middle panel),
  and $0.3830$ (right panel). In each panel, three marginally trapped surfaces
  are presented: One is the common apparent horizon, and each of
  the other two is associated with one of the two black holes.
  Compare with Fig.~\ref{BL-common-each-TTSs}.
}
\label{BL-common-each-AHs}
\end{figure}
%
%=================================%

In Sect.~\ref{section2-4} we discussed the similarity between 
(marginally) DTTSs and (marginally) trapped surfaces.
We explore the similarity further in the examples
of the Brill--Lindquist initial data.
Since the initial data are time symmetric, marginally trapped surfaces
coincide with minimal surfaces on which $k=k_1+k_2=0$ holds,
where the formulas for $k_1$ and $k_2$ are presented in
Eqs.~\eqref{k1_BL}--\eqref{D_BL} in Appendix~\ref{Appendix-A}.
Since the common marginally trapped surface is the outermost one,
it is also the (common) apparent horizon.
Although there are many works that studied marginally trapped surfaces
in the Brill--Lindquist initial data
(e.g., \cite{Cadez:1974,Bishop:1982,Gundlach:1997,Yoshino:2005}),
including the original work by Brill and Lindquist \cite{Brill:1963},
here we present the results generated by our code.

Figure~\ref{BL-common-AH-all}
shows the sections of the common apparent horizons
with the $(x,z)$-plane for
$z_0/GM=0.0$, $0.05$, $0.10$, $0.15$, $0.20$, $0.25$,
  $0.30$, $0.35$, and $0.3830$. 
For $z_0/GM\ge 0.3831$, we could not find a common apparent horizon.
For $z_0/GM=0.0$, the common apparent horizon is spherically 
symmetric with the radius $\tilde{r}/GM=1/2$, which corresponds
to the horizon radius of a Schwarzschild spacetime
in the isotropic coordinates.
As the value of $z_0/GM$ is increased, the common apparent horizon
becomes distorted. By comparing
Figs.~\ref{BL-common-all}~and~\ref{BL-common-AH-all},
a similarity between the two kinds
of surfaces can be recognized
in the response to the variation of $z_0/GM$.

Figure~\ref{BL-common-each-AHs} plots
the common apparent horizon and the two marginally trapped surfaces
each of which is associated with one of the two black holes,
for the cases $z_0/GM=0.05$ (left panel), $0.2$ (middle panel),
and $0.3830$ (right panel). 
In contrast to the marginally DTTSs, 
the two inner marginally trapped surfaces do not cross each
other. The dynamically transversely trapping regions overlap
because they cover larger domains compared to the trapped regions. 
For $z_0/GM\ge 0.3831$ no common apparent horizon
can be found, but two separate marginally trapped surfaces
can always be found.

Note that the typical size of the circumference of
the apparent horizon, $\sim$$4\pi GM$, is smaller
than that of the common marginally DTTS, $\sim$$6\pi GM$. 
Also, the parameter range $0\le z_0/GM\lesssim 0.3830$
where the common apparent horizon is present
is much smaller than the range $0\le z_0/GM\lesssim 1.1506$
where the common DTTS is present. These are because an 
apparent horizon is an indicator for a stronger
gravity region compared to a marginally DTTS.

%
%======================================%
%<<<<<<<<<<<< SECTION V  >>>>>>>>>>>>>>%
%======================================%
%
\section{Explicit examples in Majumdar--Papapetrou spacetimes}
\label{section5}

In this section we explicitly construct marginally DTTSs
in Majumdar--Papapetrou spacetimes \cite{Majumdar:1947,Papapetrou:1947}.
In Sect.~\ref{section5-1}, we explain the Majumdar--Papapetrou spacetimes
and our setups. The equation for solving a marginally DTTS
is explained in Sect.~\ref{section5-2}, and numerical results
are presented in Sect.~\ref{section5-3}. Comparison with
static TTSs is made in Sect.~\ref{section5-4}.

\subsection{Setup}
\label{section5-1}

A Majumdar--Papapetrou spacetime is a static 
electrovacuum spacetime with the metric 
\begin{equation}
  ds^2 = -U^{-2}dt^2+U^2(dx^2+dy^2+dz^2),
  \label{metric:MP}
\end{equation}
and an electromagnetic four-potential
\begin{equation}
  A_a = \frac{U^{-1}}{\sqrt{G}}(\mathrm{d}t)_a.
  \label{MP:electromaginetic-potential}
\end{equation}
We have two equations from the Einstein field equations,
which correspond to the Hamiltonian constraint
and the evolution equations, and one equation from Maxwell's equations.
These three equations are reduced to exactly the same form,
\begin{equation}
  \bar{\nabla}^2 U =0,
\end{equation}
where $\bar{\nabla}^2$ is the flat space Laplacian introduced
in Eq.~\eqref{flat-space-laplacian}. 
In general, the Majumdar--Papapetrou spacetime represents   
$N$ extremal black holes at rest. Here, we focus our attention
on the two equal-mass black holes, and 
choose the following solution to this equation,
\begin{equation}
U = 1+\frac{GM}{2\tilde{r}_+}+\frac{GM}{2\tilde{r}_-},
\end{equation}
with $\tilde{r}_\pm$ defined in Eqs.~\eqref{def:rpm},
where $M$ corresponds to the ADM mass. 
In the case of $z_0=0$, this metric represents  
an extremal Reissner-Nortstr\"om spacetime
in the isotropic coordinates and $r=0$ corresponds to
the horizon.
By contrast, in the case of $z_0>0$, 
the metric represents a spacetime
with two extremal black holes with horizons located at $(x,y,z)=(0,0,\pm z_0)$.
The two black holes are kept static because the gravitational
attraction and the electromagnetic repulsive interaction are
balanced. 

\subsection{The equation for a marginally DTTS}
\label{section5-2}

We solve for marginally DTTSs on the slice $t=\mathrm{const.}$
in this spacetime, which is time symmetric. 
Since the spatial metric in Eq.~\eqref{metric:MP} is conformally
flat, the same method for the Brill--Lindquist initial
data can be applied to this system.
The difference is that there is a nonzero contribution from
$P_{r}$ to the equation for marginally DTTSs,
Eq.~\eqref{Equation-for-a-marginally-DTTS-time-symmetric}, and
the conformal factor $\varPsi$
must be replaced by $U^{1/2}$.
The detailed forms of the equations
are presented in Appendix~\ref{Appendix-B}.

In contrast to the Brill--Lindquist case, $k_1-k_2$
changes its sign on marginally DTTSs in this spacetime.
For this reason, in solving for a marginally DTTS numerically,
we monitor the sign of $k_1-k_2$ and choose the
appropriate equation at each step of the polar angle,
$\theta=\theta_i:=i\times \Delta\theta$ ($i=0,1,...$),
in order to calculate the data
at the next step, $\theta=\theta_{i+1}$.

\subsection{Numerical results}
\label{section5-3}

%===========<FIGURE1>============%
%
\begin{figure}[tb]
\centering
\includegraphics[width=0.5\textwidth,bb=0 0 270 252]{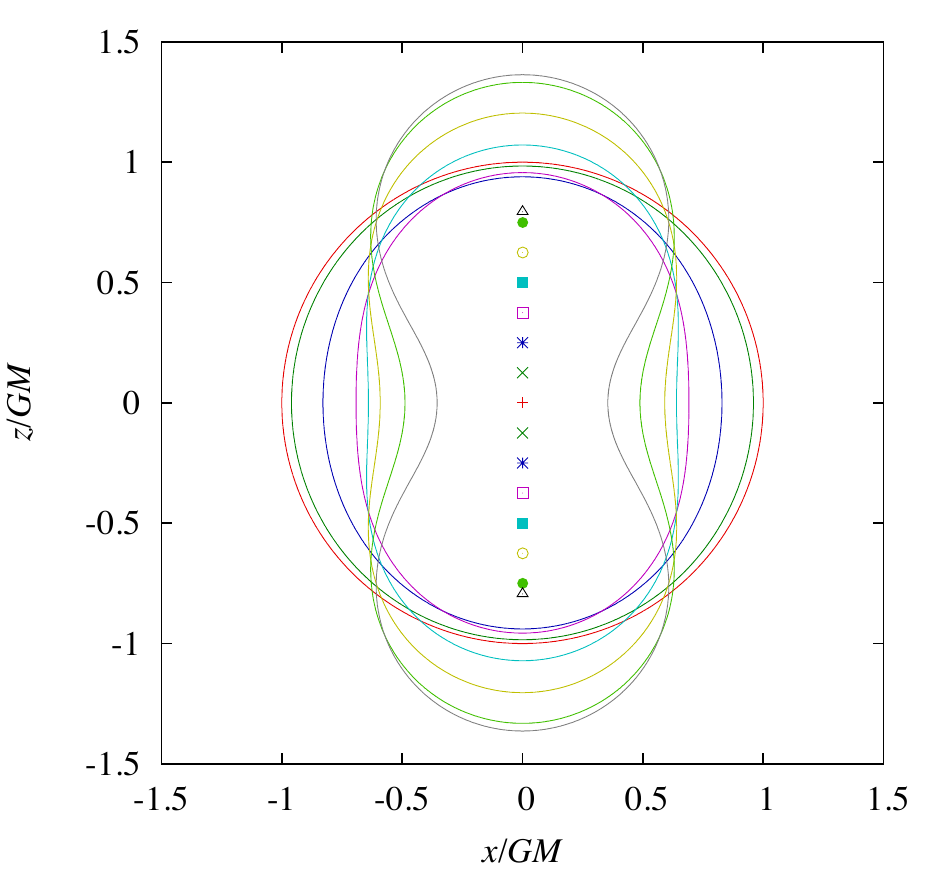}
\caption{
  Sections of common marginally DTTSs with the $(x,z)$-plane
  in the Majumdar--Papapetrou spacetimes 
  for $z_0/GM=0.0$, $0.125$, $0.25$, $0.375$, $0.5$, $0.625$, $0.75$,
  and $0.79353$.
  For $z_0/GM\ge 0.79354$, no common marginally DTTS is present.  
}
\label{MP-common-all}
\end{figure}
%
%=================================%

%===========<FIGURE1>============%
%
\begin{figure}[tb]
\centering
\includegraphics[width=0.96\textwidth,bb=0 0 454 169]{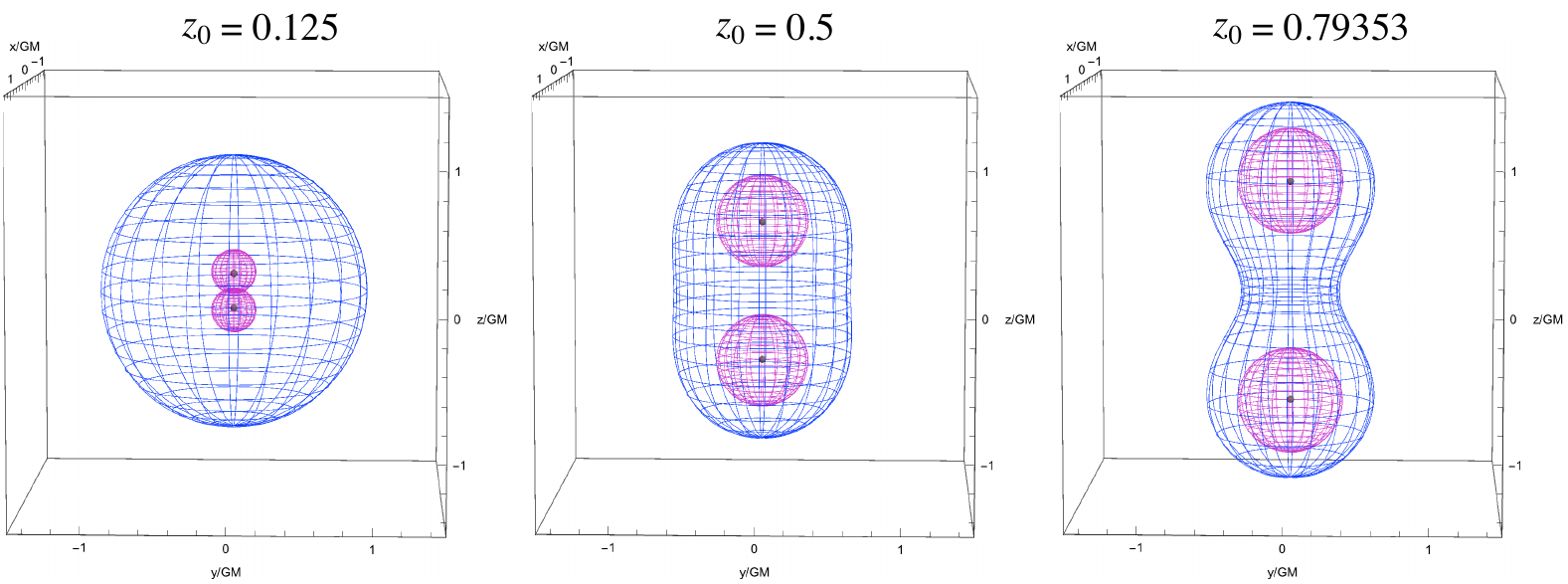}
\caption{
  3D plots of marginally DTTSs in the Majumdar--Papapetrou spacetime
  for $z_0/GM=0.125$ (left panel), $0.5$ (middle panel),
  and $0.79353$ (right panel). In each panel, three marginally DTTSs
  are presented: One is the common DTTS, and each of
  the other two DTTSs encloses only one black hole.   
}
\label{MP-common-each-TTSs}
\end{figure}
%
%=================================%

We now show the numerical results. Figure~\ref{MP-common-all}
shows the sections
of the common marginally DTTSs with the $(x,z)$-plane 
in the Majumdar--Papapetrou spacetimes
for $z_0/GM=0.0$, $0.125$, $0.25$, $0.375$, $0.5$, $0.625$, $0.75$, and $0.79353$.
For $z_0/GM=0.0$, the common marginally DTTS is spherically
symmetric with the radius $\tilde{r}/GM=1$. This corresponds
to the radius of the photon sphere of an extremal Reissner-Nordstr\"om
spacetime in the isotropic coordinates.
As the value of $z_0/GM$ is increased, the common DTTS becomes distorted,
and it exists up to $z_0/GM\approx 0.79353$.
Similarly to the Brill--Lindquist case, 
for the range $0.647\lesssim z_0/GM\lesssim 0.79353$
we could obtain two solutions that correspond to the outer
and inner boundaries of 
a common dynamically transversely
trapping region. Around $z_0/GM\approx 0.79353$, the inner
and outer boundaries degenerate, and  
the common dynamically transversely trapping region vanishes
as $z_0/GM$ is further increased.

Figure~\ref{MP-common-each-TTSs} shows 3D plots of the marginally DTTSs
for the cases $z_0/GM=0.125$ (left panel), $0.5$ (middle panel),
and $0.79353$ (right panel). In this figure,
we plot both the common marginally DTTS and the
marginally DTTSs, each of which surrounds only one of the two black holes.
As shown in the left panel,
the two inner marginally DTTSs cross with each
other for small $z_0/GM$, and the two dynamically
transversely trapping regions overlap. 
As $z_0/GM$ is increased, the two dynamically
transversely trapping regions become separate
as shown in the middle and right panels.
Although no common DTTS 
can be found for $z_0/GM\ge 0.79354$, 
the two separate DTTSs can always be found.

%===========<FIGURE1>============%
%
\begin{figure}[tb]
\centering
\includegraphics[width=0.9\textwidth,bb=0 0 464 163]{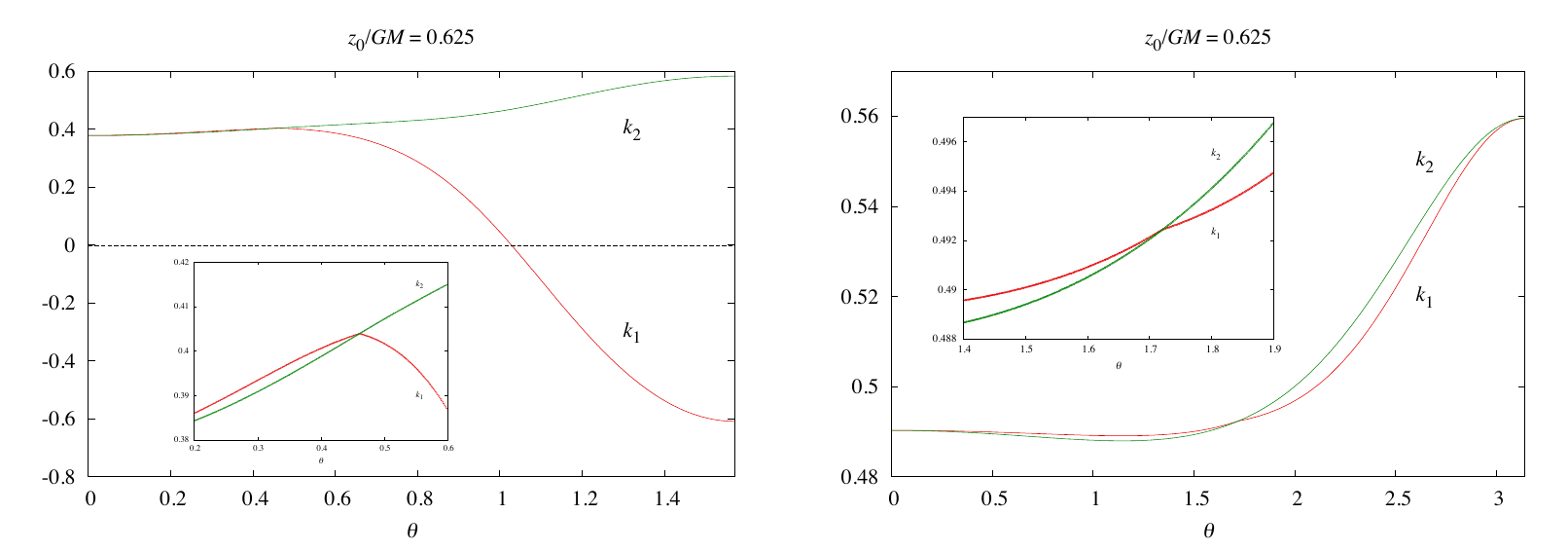}
\caption{
  Behavior of $k_1$ and $k_2$ on marginally DTTSs as functions of the polar
  angle $\theta$ in the Majumdar--Papapetrou case. Left panel:
  $k_1$ and $k_2$ for the common marginally DTTS in the case $z_0/GM=0.625$.
  Right panel: $k_1$ and $k_2$ for the marginally DTTS that encloses
  only the upper black hole in the case $z_0/GM=0.625$.
  In both panels, $k_1-k_2$ changes its sign from positive to negative
  as $\theta$ is increased. In each of the two panels, 
  the inset enlarges the neighborhood of the point
  where $k_1-k_2$ changes its sign. 
}
\label{k1_k2_MP}
\end{figure}
%
%=================================%

Figure~\ref{k1_k2_MP} shows the values of $k_1$ and $k_2$
as functions of the polar angle $\theta$
for a common DTTS 
(left panel) and for a DTTS that surrounds only the upper
black hole (right panel) for the case $z_0/GM=0.625$.
In each panel, $k_1-k_2$ changes its sign at some $\theta=\theta_{\rm c}$,
and 
the equations are changed in the domains $\theta\le \theta_{\rm c}$
and $\theta_{\rm c}\le \theta$, accordingly.
The curve for $k_1$ is bent at $\theta=\theta_{\rm c}$, 
because, due to
the change of the equations, the third derivative of $h(\theta)$
is discontinuous at $\theta=\theta_{\rm c}$.
Since $k_1$ depends on $h^{\prime\prime}$ as presented in Eqs.~\eqref{k1_MP}
and \eqref{C_MP},
the derivative of $k_1$ becomes discontinuous, although the curve
of $k_1$ itself is continuous. By contrast,
since $k_2$ does not depend on $h^{\prime\prime}$
as presented in Eqs.~\eqref{k2_MP} and \eqref{D_MP},
the curve of $k_2$ is not bent at $\theta=\theta_{\rm c}$.
These results indicate that the obtained marginal DTTSs
are of differentiability class $C^2$ in this case.

\subsection{Comparison with static TTSs}
\label{section5-4}

Since the Majumdar--Papapetrou spacetime is a static spacetime,
the static TTSs defined in our previous paper \cite{Yoshino:2017}
can also be studied. Here, we study
common static TTSs that enclose both black holes
and compare them with common DTTSs.

A static timelike surface
$S$ is said to be a static TTS if and only if
arbitrary photons emitted from arbitrary points on $S$
in the tangential direction to $S$ propagate on
$S$ or toward the inside of $S$. This condition,
hereafter the static TTS condition, is expressed as
$\bar{K}_{ab}k^ak^b\le 0$, where $k^a$ is
arbitrary null tangent vectors to $S$.
The static TTS condition is rewritten as
\begin{equation}
  \mathrm{max}(k_1, k_2)\le \frac{{}^{(3)}\pounds_{\hat{r}}\alpha_{\rm s}}{\alpha_{\rm s}},
  \label{Static-TTS-condition}
\end{equation}
on a static slice $t=\mathrm{const.}$,
where $\alpha_{\rm s}:=\sqrt{-\xi_a\xi^a}=U^{-1}$ is the
lapse function associated with the timelike Killing vector $\xi^a$. 
Unlike the marginally transversely trapping condition in the DTTS case,
we do not require the existence of a photon that
marginally satisfies the condition in Eq.~\eqref{Static-TTS-condition}.

%===========<FIGURE1>============%
%
\begin{figure}[tb]
\centering
\includegraphics[width=0.96\textwidth,bb=0 0 444 149]{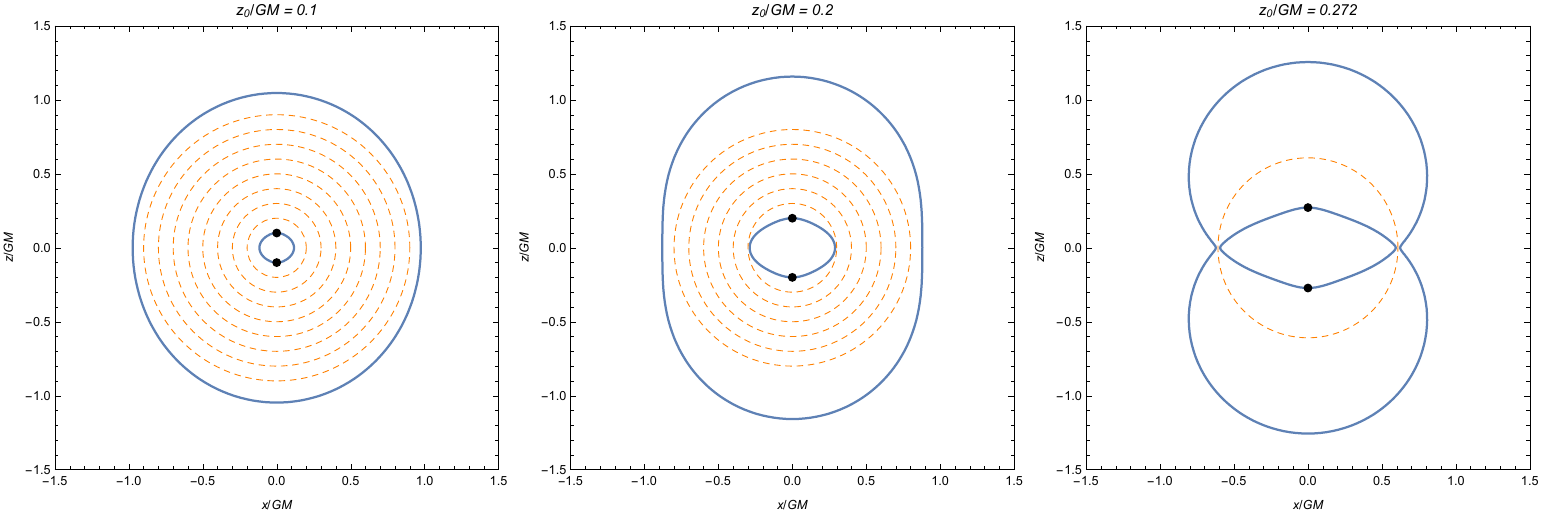}
\caption{
  Examples of common static TTSs in the Majumdar--Papapetrou spacetime
  for $z_0/GM=0.1$ (left panel), $0.2$ (middle panel),
  and $0.272$ (right panel). Two solid curves indicate
  the contours of $U+2\tilde{r}U_{,\tilde{r}}= 0$, and
  $\tilde{r}=\mathrm{const.}$ surfaces between the two curves
  are static TTSs, as some of them are indicated by dashed circles.
}
\label{MP-common-static-TTSs}
\end{figure}
%
%=================================%

First, let us consider the condition that a surface $\tilde{r}=\mathrm{const.}$
becomes a common static TTS. The formulas for $k_1$ and $k_2$ are
given in Eqs.~\eqref{k1_MP}--\eqref{D_MP} in Appendix~\ref{Appendix-B}.
Since $k_1=k_2=U^{-1}(1/\tilde{r}+U_{,\tilde{r}}/U)$
holds for a surface $\tilde{r}=\mathrm{const.}$,
the static TTS condition is reduced to
\begin{equation}
  U+2\tilde{r}U_{,\tilde{r}}\le 0.
  \label{Static-TTS-condition-MP}
\end{equation}
Figure~\ref{MP-common-static-TTSs} shows the contours of
$U+2\tilde{r}U_{,\tilde{r}}= 0$ in the cases of $z_0/GM=0.1$, $0.2$, and $0.272$.
All of the surfaces $\tilde{r}=\mathrm{const.}$ between the two contours 
are common static TTSs, and some of them are indicated by dashed circles.
The surfaces $\tilde{r}=\mathrm{const.}$ can become TTSs only
if $z_0$ is within the range $0\le z_0/GM\le \sqrt{2/27}\approx 0.2722$.
For $\sqrt{2/27}< z_0/GM$, the contour surfaces are reconnected 
and a surface $\tilde{r}=\mathrm{const.}$ cannot cross the equatorial plane 
without violating 
the static TTS condition of Eq.~\eqref{Static-TTS-condition-MP}.

Next, let us consider the general case where a common static TTS
is given by $\tilde{r}=h(\tilde{\theta})$. As a necessary condition,
a common static TTS must satisfy $k_2\le {}^{(3)}\pounds_{\hat{r}}\alpha_{\rm s}/\alpha_{\rm s}$
on the equatorial plane. Since $\cot\tilde{\theta}=U_{,\tilde{\theta}}=0$
at $\tilde{\theta}=\pi/2$ in Eqs.~\eqref{k2_MP} and \eqref{D_MP},
this necessary condition is reduced to
\begin{equation}
  \left.(U+2\tilde{r}U_{,\tilde{r}})\right|_{\tilde{r}=h(\pi/2)}\le 0,
  \label{Static-TTS-condition-MP-equatorial-plane}
\end{equation}
which is the same as the static TTS condition in 
Eq.~\eqref{Static-TTS-condition-MP} for $\tilde{r}=\mathrm{const.}$
surfaces. Therefore, we find the following: if there is no
radius in the equatorial plane that satisfies
the condition in Eq.~\eqref{Static-TTS-condition-MP-equatorial-plane},
there is no common static TTS. Conversely, if there is a radius
in the equatorial plane that satisfies
the condition in Eq.~\eqref{Static-TTS-condition-MP-equatorial-plane},
common static TTSs are present (for example, the surface $\tilde{r}=\mathrm{const.}$).
The condition for the existence of common static TTSs is determined
only on the equatorial plane.

Physically, this result is directly related to the (non)existence
of circular orbits of photons on the equatorial plane.
As studied in Ref.~\cite{Wunsch:2013}, two circular orbits of photons
are located at the radius at which   
the equality in the
condition of Eq.~\eqref{Static-TTS-condition-MP-equatorial-plane} holds. 
Therefore, if two circular orbits of photons exist, common static
TTSs are present because they can cross the equatorial plane at
the radius between the 
two circular orbits of photons.
If there are no circular orbits of photons, common static TTSs
cannot cross the equatorial plane anymore.

Notice that the parameter range $0\le z_0/GM\lesssim 0.2722$ for the existence
of common static TTSs is much smaller than the range
$0\le z_0/GM\lesssim 0.7935$
for the existence of common DTTSs. Such difference occurs 
because the timelike surface $S$ must not change in 
time in the case of a static TTS, while $S$ 
is flexible and can change its shape
in time in the case of a DTTS.
Due to this flexibility of $S$ in the case of a DTTS,
even if an area element of $\sigma_t$ (defined in
Sect.~\ref{section2-3-1} as a $t=\mathrm{const.}$ slice of $S$) 
expands in the $\phi$ 
direction in time, the 
accelerated contraction condition is not necessarily violated, because
${}^{(3)}\bar{\pounds}_n\bar{k}\le 0$ is satisfied if 
the area element contracts 
in the $\theta$ direction sufficiently rapidly.
Therefore, the concept of a DTTS is fairly different from that
of a static TTS.

%
%======================================%
%<<<<<<<<<<<< SECTION VI  >>>>>>>>>>>>>>%
%======================================%
%
\section{Penrose-like inequality}
\label{section6}

In the following two sections, we focus on
general properties of DTTSs.
In this section we prove that DTTSs satisfy
the Penrose-like inequality in Eq.~\eqref{Penroselike-inequality}
under certain conditions. Before doing this,
it is useful to review the Penrose inequality, Eq.~\eqref{Penrose-inequality}.

The Penrose inequality \eqref{Penrose-inequality} is conjectured
to be satisfied by an apparent horizon by the following argument.
If the cosmic censorship hypothesis holds, there is an event
horizon outside the apparent horizon, and the area of the event
horizon $A_{\rm EH}$ is expected to be equal to or larger than that of the
apparent horizon, i.e.  $A_{\rm AH}\le A_{\rm EH}$.
Due to the area theorem, the event horizon has the area $A_{\rm EH}^{\rm (f)}$,
larger than $A_{\rm EH}$,
after the system settles down to a stationary 
state described by a Kerr black hole, i.e. $A_{\rm EH}\le A_{\rm EH}^{\rm (f)}$.
Furthermore, the relation 
$A_{\rm EH}^{\rm (f)}\le 4\pi (2GM)^2$ is expected to be
satisfied due to the positivity of
the radiated energy of gravitational waves.
This leads to the Penrose conjecture,
the inequality in Eq.~\eqref{Penrose-inequality}.
Note that a counterexample to the Penrose conjecture 
has been constructed in Ref.~\cite{BenDov:2004} by cutting and gluing
Schwarzschild spacetimes and Friedmann universes in a complex manner.
In this system,
the Penrose inequality is not satisfied because 
the inequality $A_{\rm AH}\le A_{\rm EH}$ in the above discussion is violated.
However, it is possible to reformulate the Penrose conjecture
to be consistent with the above counterexample as argued in Ref.~\cite{BenDov:2004}.

Although the Penrose conjecture remains an open problem,
the Riemannian Penrose inequality, which is a variant of the original
Penrose inequality, has been proved. If an asymptotically
flat three-dimensional space $\Sigma$ with nonnegative scalar curvature
possesses an outermost minimal surface with the area $A_{\rm min}$,
the Riemannian Penrose inequality asserts
\begin{equation}
  A_{\rm min}\le 4\pi (2GM)^2.
  \label{Riemannian-Penrose-inequality}
\end{equation}
As a corollary, the Riemannian Penrose inequality implies
that the original Penrose inequality is satisfied
by an apparent horizon in time-symmetric initial data.
There are two methods to prove the Riemannian Penrose inequality:
the method by the inverse mean curvature flow \cite{Wald:1977,Huisken:2001}
and Bray's conformal flow \cite{Bray:2001}.
Of these, the method by the inverse mean curvature
flow is directly used in this section.

The inverse mean curvature flow is an example of a geometric
flow of hypersurfaces of a Riemannian manifold.
Let us consider a flow of two-dimensional hypersurfaces $\sigma(r)$
with spherical topology in $\Sigma$, each of which is labeled by
the radial coordinate $r$. 
If the lapse function $\varphi$ satisfies $\varphi=1/k$, this flow
is said to be the inverse mean curvature flow.
For each of the surfaces of the flow, Geroch's quasilocal mass
is defined by 
\begin{equation}
E(r) = \frac{A^{1/2}}{64\pi^{3/2}G}\left(16\pi-\int_{\sigma(r)}k^2dA\right),
\end{equation}
where $A$ and $dA$ are the area and the area element of $\sigma(r)$.  
Geroch's quasilocal mass is checked to coincide with the ADM mass
at spacelike infinity, $r\to \infty$.
Furthermore, for a space with nonnegative Ricci scalar,
Geroch's quasilocal mass is proved to satisfy 
\begin{equation}
  \frac{dE}{dr} \ge 0,
\end{equation}
which is called the Geroch monotonicity \cite{Geroch:1973}.
Due to the Geroch monotonicity,
Geroch's quasilocal mass for the surface $\sigma_0=\sigma(0)$, i.e. $r=0$,
satisfies $E(0)\le M$. 
If the surface $\sigma_0$ is an infinitesimal limit of an $S^2$ surface,
we have
\begin{equation}
E(0)=0\le M,
\end{equation}
which proves the positive energy theorem \cite{Geroch:1973}. If
the surface $\sigma_0$ is a minimal surface on which $k=0$ is satisfied,
we have
\begin{equation}
E(0) = \frac{A_{\rm min}^{1/2}}{4\pi^{1/2}G}\le M.
\end{equation}
This implies the Riemannian Penrose inequality of Eq.~\eqref{Riemannian-Penrose-inequality}
as pointed out in Ref.~\cite{Wald:1977}. 
Note that singularities appear in the inverse mean curvature flow
in general. However, Huisken and Ilmanen \cite{Huisken:2001}
showed that it is possible to
introduce a weak solution to the inverse mean curvature flow
without breaking the Geroch monotonicity, and thus gave a 
complete proof of the Riemannian Penrose inequality.

In order to prove that DTTSs satisfy the Penrose-like inequality
of Eq.~\eqref{Penroselike-inequality},
we will show that a DTTS satisfies
\begin{equation}
  \int_{\sigma_0}k^2 dA\le \frac{16}{3}\pi
  \label{condition-for-integral_k2_dA}
\end{equation}
under certain conditions. Then, the Geroch monotonicity
implies that
\begin{equation}
\frac{A_{0}^{1/2}}{6\pi^{1/2}G}\le E(0)\le M,
\end{equation}
which is equivalent to the Penrose-like inequality
in Eq.~\eqref{Penroselike-inequality}.
We discuss the cases of the time-symmetric initial data
and the momentarily stationary axisymmetric initial data
one by one.

\subsection{Time-symmetric initial data}

Suppose we have a DTTS $\sigma_0$ in time-symmetric initial data $\Sigma$.
The DTTS $\sigma_0$ satisfies the formula for ${}^{(3)}\bar{\pounds}_{n}\bar{k}$
given by Eq.~\eqref{Lie-derivative-bark-time-symmetric}.
In this equation, 
the inequality ${}^{(3)}\bar{\pounds}_{n}\bar{k} \le 0$
holds from the accelerated contraction condition of Eq.~\eqref{accelerated-contraction-condition}. 
We assume the radial pressure to be nonpositive, $P_{r}\le 0$.
Furthermore, if the convexity $k_{\rm S}\ge 0$ is assumed for 
the DTTS, we have
\begin{equation}
  2kk_{\rm L}+k^2-k_{ab}k^{ab}
  =
  \frac{3}{2}k^2 + \frac12(k_{\rm L}+3k_{\rm S})(k_{\rm L}-k_{\rm S}) \ge \frac{3}{2}k^2.
  \label{k2-inequality-time-symmetric}
\end{equation}
Then, we have the inequality, ${}^{(2)}R \ge (3/2)k^2$ 
and integration over the surface $\sigma_0$ gives 
\begin{equation}
\int_{\sigma_0}k^2dA\le \frac{2}{3}\int_{\sigma_0}{}^{(2)}RdA.
\end{equation}
If there is a point at which $k>0$ holds, the Gauss--Bonnet theorem
tells us that $\sigma_0$ has topology $S^2$ and satisfies
$\int_{\sigma_0}{}^{(2)}RdA=8\pi$. Therefore,
the inequality in Eq.~\eqref{condition-for-integral_k2_dA} is satisfied,
and thus we have shown the following:

%===========<THEOREM>============%
%
\begin{theorem}
  A convex DTTS, $\sigma_0$, in time-symmetric, asymptotically
  flat initial data 
  has topology $S^2$ and  
  satisfies the Penrose-like inequality
  $A_0\le 4\pi (3GM)^2$ if $P_r\le 0$ holds on $S_0$,
  $k>0$ at least at one point on $S_0$, and ${}^{(3)}R$
  is nonnegative (i.e. the energy density $\rho\ge 0$)
  in the outside region.
  \label{Theorem:Penroselike-inequality-time-symmetric}
\end{theorem}
%
%=================================%

Note that, by virtue of the inequality in Eq.~\eqref{k2-inequality-time-symmetric},
Theorem~\ref{Theorem:Penroselike-inequality-time-symmetric}
also holds for a nonconvex DTTS as well if $k_{\rm S}$ 
is within the range $0>k_{\rm S}\ge -k_{\rm L}/3$.
Although nonpositive radial pressure $P_{r}\le 0$ may seem strange,
it is not very unrealistic because 
it is satisfied if the spacetime is vacuum around $\sigma_0$, and 
furthermore, the radial pressure due to electromagnetic fields 
is negative on spherical 
surfaces in a Reissner--Nordstr\"om spacetime.

\subsection{Momentarily stationary axisymmetric initial data}

In the case of the momentarily stationary axisymmetric initial data,
the formula for ${}^{(3)}\bar{\pounds}_{n}\bar{k}$
is given by
Eq.~\eqref{Lie-derivative-bark-momentarily-stationary-axisymmetric}.
Similarly to the time-symmetric case, the accelerated contraction
condition implies that ${}^{(3)}\bar{\pounds}_{n}\bar{k}\le 0$, and $P_r\le 0$ is assumed.
Assuming $\sigma_0$ to be a convex surface with $k_{\rm S}\ge 0$, 
the relation $2k{}^{(3)}\pounds_{\hat{r}}\alpha/\alpha\ge 2kk_{\rm L}$ holds
from the inequality 
in Eq.~\eqref{inequality-Lr-alpha-momentarily-stationary-axisymmetric}
and $k\ge 0$, and thus we have 
\begin{equation}
  2k\frac{{}^{(3)}\pounds_{\hat{r}}\alpha}{\alpha}+k^2-k_{ab}k^{ab}
  \ge \frac{3}{2}k^2,
  \label{k2-inequality-momentarily-stationary-axisymmetric}
\end{equation}
by the same calculation as Eq.~\eqref{k2-inequality-time-symmetric}.
Since it is difficult to control the sign of $({}^{(3)}\pounds_{\hat{r}}\tilde{\omega})^2-(\mathcal{D}\tilde{\omega})^2$,
we simply assume that $({}^{(3)}\pounds_{\hat{r}}\tilde{\omega})^2\ge (\mathcal{D}\tilde{\omega})^2$.
In other words, we choose surfaces  
$\sigma_0$ so that this condition is satisfied.
Then, we have the inequality ${}^{(2)}R \ge (3/2)k^2$,
and with the same argument as the time-symmetric case,
the inequality in Eq.~\eqref{condition-for-integral_k2_dA} is satisfied.
Therefore, we have shown the following:

%===========<THEOREM>============%
%
\begin{theorem}
  An axisymmetric convex DTTS 
  $\sigma_0$ in momentarily stationary axisymmetric initial data  
  has topology $S^2$  and
  satisfies the Penrose-like inequality
  $A_0\le 4\pi (3GM)^2$ if $P_r\le 0$ and
\begin{equation}
  ({}^{(3)}\pounds_{\hat{r}}\tilde{\omega})^2
  \ge
  (\mathcal{D}\tilde{\omega})^2
  \label{AxiStationary-Penrose-Assumption}
\end{equation}
  hold on $\sigma_0$, $k>0$ at least at one point on $\sigma_0$, and ${}^{(3)}R$
  is nonnegative in the outside region.
\label{Theorem:Penroselike-inequality-momentarily-stationary-axisymmetric}
\end{theorem}
%
%=================================%
Similarly to the time-symmetric case,
Theorem~\ref{Theorem:Penroselike-inequality-momentarily-stationary-axisymmetric}
also holds for a nonconvex DTTS as well if $k_{\rm S}$ 
is within the range $0>k_{\rm S}\ge -k_{\rm L}/3$
due to the inequalities in Eqs.~\eqref{k2-inequality-time-symmetric}~and~\eqref{k2-inequality-momentarily-stationary-axisymmetric}.

%
%======================================%
%<<<<<<<<<<<< SECTION VII  >>>>>>>>>>>>>>%
%======================================%
%
\section{Connection to loosely trapped surfaces}
\label{section7}

Lastly, we study the connection between
DTTSs and LTSs. The LTS is defined in our previous paper \cite{Shiromizu:2017}
as a surface on which $k>0$ and ${}^{(3)}\pounds_{\hat{r}}k\ge 0$
are satisfied in a flow of two-dimensional closed surfaces
in a spacelike hypersurface $\Sigma$.
In fact, such surfaces are located only between the horizon $r=2GM$
and the photon surface $r=3GM$
in the Schwarzschild case. It was also proved in Ref.~\cite{Shiromizu:2017}
that an LTS satisfies the inequality in Eq.~\eqref{condition-for-integral_k2_dA},
and hence satisfies the Penrose-like inequality, Eq.~\eqref{Penroselike-inequality},
if $\Sigma$ has a nonnegative Ricci
scalar, ${}^{(3)}R\ge 0$. In Ref.~\cite{Shiromizu:2017},
clarifying the relation of LTSs to the behavior
of photons was left as a remaining problem, and we show here
the fact that a DTTS is an LTS at the same time
under certain conditions.

We derive a formula that relates ${}^{(3)}\bar{\pounds}_n\bar{k}$
and ${}^{(3)}\pounds_{\hat{r}}k$.
The trace of the Ricci equation on $\sigma_0$ as a hypersurface
in $\Sigma$ is
\begin{equation}
  {}^{(3)}{\pounds}_{\hat{r}}{k}
  =-{}^{(3)}{R}_{ab}\hat{r}^a\hat{r}^b
  -k_{ab}k^{ab}-\frac{1}{\varphi}\mathcal{D}^2\varphi,
  \label{trace-Ricci-sigma0-in-Sigma}
\end{equation}
where ${}^{(3)}{R}_{ab}$ is the Ricci tensor associated with
the metric $q_{ab}$ induced on $\Sigma$. 
This equation is rewritten with the double trace of the Gauss equation
of $\sigma_0$ in $\Sigma$,
\begin{equation}
  {}^{(2)}R = {}^{(3)}{R} - 2{}^{(3)}{R}_{ab}\hat{r}^a\hat{r}^b
  +k^2-k_{ab}k^{ab},
\end{equation}
and the 
double trace of the Gauss equation on $\Sigma$ in the spacetime $\mathcal{M}$,
\begin{equation}
{}^{(3)}R = 2G_{ab}n^an^b-K^2+K_{ab}K^{ab}.
\end{equation}
The result is 
\begin{equation}
  {}^{(3)}{\pounds}_{\hat{r}}k = \frac12{}^{(2)}R
  -8\pi G\rho
  +\frac12\left(K^2-K_{ab}K^{ab}-k^2-k_{ab}k^{ab}\right)
  -\frac{1}{\varphi}\mathcal{D}^2\varphi,
  \label{Lie-derivative-of-k-rewritten}
\end{equation}
where the Einstein field equations are assumed.
Adding Eqs.~\eqref{Lie-derivative-of-bark-rewritten} 
and \eqref{Lie-derivative-of-k-rewritten} and rewriting
with the decomposed forms
in Eqs.~\eqref{barKab-decomposition}~and~\eqref{Kab-decomposition}
of $\bar{K}_{ab}$ and $K_{ab}$, respectively, 
we have
\begin{multline}
   {}^{(3)}{\pounds}_{\hat{r}}k 
  =-{}^{(3)}\bar{\pounds}_{n}\bar{k} -8\pi G(\rho+P_{r})
  + k\frac{{}^{(3)}\pounds_{\hat{r}}\alpha}{\alpha}
  -k_{ab}k^{ab}
  -\frac{1}{\varphi}\mathcal{D}^2\varphi
  -\bar{k}_{ab}\bar{k}^{ab}
  \\
  + \bar{k}\frac{{}^{(3)}\bar{\pounds}_{n}\varphi}{\varphi}
  +\frac{1}{\alpha}\mathcal{D}^2\alpha. 
\label{pound_n-bark-plus-pound_hatr-k}
\end{multline}

We apply this formula to a DTTS $\sigma_0$. The last two terms vanish
because $\alpha=\mathrm{const.}$
and $\bar{k}=0$ are required.
Let us evaluate the sign
of each term on the right-hand side of
Eq.~\eqref{pound_n-bark-plus-pound_hatr-k}.
The first term is nonnegative, 
$-{}^{(3)}\bar{\pounds}_{n}\bar{k} \ge 0$, from the
accelerated contraction condition of Eq.~\eqref{accelerated-contraction-condition}.
The second term is nonpositive as long as
the dominant energy condition holds. For this reason, 
we require $\rho+P_r=0$ in order to make the left-hand side
positive definite.
As for the third and fourth terms,
for both of the time-symmetric and
momentarily stationary axisymmetric initial data,
we have
\begin{equation}
k\frac{{}^{(3)}\pounds_{\hat{r}}\alpha}{\alpha}
-k_{ab}k^{ab}
\ge kk_{\rm L}-k_{ab}k^{ab}
=k_{\rm S}(k_{\rm L}-k_{\rm S})\ge 0
\end{equation}
for a convex DTTS with $k_{\rm S}\ge 0$, 
using the marginally transversely trapping conditions 
in the time-symmetric 
and momentarily stationary axisymmetric cases,
Eqs.~\eqref{marginally-transversely-trapping-condition-time-symmetric}~and~\eqref{marginally-transversely-trapping-condition-momentarily-stationary-axisymmetric-case}, respectively.
We set the fifth term to be zero by choosing $\varphi=\mathrm{const.}$,
since $\sigma_0$ becomes an LTS if ${}^{(3)}\pounds_{\hat{r}}k\ge 0$ is satisfied 
at least for one chice of $\varphi$. We discuss
the sixth term $-\bar{k}_{ab}\bar{k}^{ab}$ in the time-symmetric and momentarily
stationary axisymmetric cases separately.

\subsection{Time-symmetric initial data}

Since $\bar{k}_{ab}=0$ holds for time-symmetric initial data,
${}^{(3)}{\pounds}_{\hat{r}}{k} \ge 0$ is guaranteed
only with the conditions discussed above.
Therefore, we have found the following proposition:

%===========<PROPOSITION>============%
%
\begin{proposition}
A convex DTTS $\sigma_0$ with $k>0$ in time-symmetric initial data 
is an LTS as well if $\rho+P_r=0$ is satisfied on $\sigma_0$.
\label{proposition-1}
\end{proposition}
%
%=================================%

\subsection{Momentarily stationary axisymmetric initial data}

For momentarily stationary axisymmetric initial data, $\bar{k}_{ab}$
is given by Eq.~\eqref{barkab:momentarily-stationary-axisymmetric},
and the sixth term becomes
\begin{equation}
  -{\bar{k}_{ab}}{\bar{k}^{ab}}
  =-\frac{\phi_a\phi^a}{2\tilde{\alpha}^2}(\mathcal{D}\tilde{\omega})^2 \le 0.
\end{equation}
In order to make this term vanish, we have to
require $\mathcal{D}_a\tilde{\omega}=0$. This means that
we limit our discussion to 
$\tilde{\omega}=\mathrm{const.}$ surfaces:

%===========<PROPOSITION>============%
%
\begin{proposition}
  If a contour surface of $\tilde{\omega}$
  in momentarily stationary axisymmetric initial data 
  is a convex DTTS on which $\rho+P_r=0$ is satisfied,
  it is an LTS as well.
  \label{proposition-2}
\end{proposition}
%
%=================================%

%
%======================================%
%<<<<<<<<<<<< SECTION VIII  >>>>>>>>>>>>>>%
%======================================%
%
\section{Summary and discussion}
\label{section8}

In this paper, we have defined a (marginally) dynamical transversely
trapping surface (DTTS) as an extended concept
of the photon sphere,
intending to provide a new theoretical tool to advance our 
understanding of the properties of dynamically evolving spacetimes
with strong gravity regions. 
The definition is given in Sect.~\ref{section2-2}.
Intuitively, a DTTS is a two-dimensional closed surface
on a spacelike hypersurface $\Sigma$
such that 
photons emitted from it in the transverse directions experience
accelerated contraction during the propagation affected by strong gravity.
The key quantity is ${}^{(3)}\bar{\pounds}_{\bar{n}}\bar{k}$,
which is required to be
nonpositive by the accelerated contraction condition.
This quantity is found
from the study on a photon surface in a Schwarzschild spacetime
(Sect.~\ref{section2-1}).
As discussed in Sects.~\ref{section2-3-3} and \ref{section5-4}, 
the concept of DTTSs is different from that of static/stationary
TTSs, which was proposed as surfaces to characterize strong gravity regions
in static/stationary spacetimes
in our previous paper \cite{Yoshino:2017}.
These two concepts must be distinguished.

We have prepared the method of solving for
a marginally DTTS in the time-symmetric
initial data and the momentarily stationary axisymmetric initial data
(Sect.~\ref{section3}). By constructing numerical
solutions explicitly for systems of two equal-mass black holes
in the Brill--Lindquist initial data (Sect.~\ref{section4})
and in the Majumdar--Papapetrou spacetimes (Sect.~\ref{section5}),
we have shown that a marginally DTTS is a well-defined concept. 
Extending the method to other configurations is necessary,
and we plan to study this issue in a forthcoming paper.

Marginally DTTSs are defined with the intention to make them
analogous to marginally trapped surfaces, and we 
have stressed various aspects of such similarity.
Both surfaces are determined on a given spacelike hypersurface $\Sigma$,
and have similar gauge-independent 
and -dependent features as discussed in Sect.~\ref{section2-4}.
In the Brill--Lindquist initial data,
their shapes and dependence on the system parameter $z_0/GM$
show qualitatively similar behavior
(Sects.~\ref{section4-3} and \ref{section4-4}).
Furthermore, we have shown that the area of a DTTS satisfies
the Penrose-like inequality in Eq.~\eqref{Penroselike-inequality}
under certain conditions in Sect.~\ref{section6},
similarly to 
the area of a marginally trapped surface being conjectured
(and partly proved) 
to satisfy the Penrose inequality in Eq.~\eqref{Penrose-inequality}. 
In addition, in Sect.~\ref{section7} 
we have discussed the fact that
DTTSs are connected to LTSs 
proposed in our previous paper \cite{Shiromizu:2017}
under some situations.

Further similarity between marginally DTTSs and marginally trapped surfaces
could be explored. For example, as the condition for
the formation of apparent horizons, the hoop conjecture \cite{Thorne:1972}
has been proposed: 
``Black holes with horizons form when and only when a mass $M$
gets compacted into a region whose circumference in every direction
is bounded by $C\lesssim 2\pi (2GM)$.''
Although no solid proof has been found up to now, this conjecture is
checked to be satisfied in various situations (e.g. \cite{Chiba:1994,Yoshino:2001,Yoshino:2007}). 
One of the implications of this conjecture is that
the apparent horizon cannot become arbitrarily long in one direction.
The analogous condition, $C\lesssim 2\pi (3GM)$, may be expected to hold
for the formation of marginally DTTSs.
We are planning to study this issue in future.

The concept of a trapped surface
has become important through
the singularity theorems (see pp.~239--241 of Ref.~\cite{Wald}).
Assuming cosmic censorship,
the existence of a trapped surface 
implies the presence of an event horizon outside.
Therefore, the existence of a trapped surface strongly
restricts the global property of a spacetime.
Does the existence of a DTTS restrict the global structure
of a spacetime as well?
Unfortunately, this is unlikely under broad assumptions
because studies on a spherically symmetric barotropic star
do not necessarily exclude a star with radius
smaller than $3GM$ as general
arguments \cite{Buchdahl:1959,Barraco:2002,Fujisawa:2015}.
However, a detailed numerical study indicates that the radii of 
spherical polytropic stars cannot be smaller than $3GM$ \cite{Saida:2015},
and therefore, for restricted situations, the presence of a
DTTS may result in the formation of an event horizon.
This issue is worth challenging. 
Note that, since photons in the definition of a DTTS are emitted
in transverse directions, a collection of corresponding null geodesics
is not an ordinary null geodesic congruence. For this reason, 
a new technology to handle the propagation of such photons
should be required.
Related to this issue, a ``wandering set'' was recently proposed 
as an extension of a photon sphere, $r=3GM$ in the Schwarzschild case, 
from the global point of view \cite{Siino:2019}. 
Since a wandering set would be 
an analogous concept to an event horizon as a generalization
of the horizon, $r=2GM$ in the Schwarzschild case, 
the concepts of DTTSs and 
a wandering set may be related to each other
like trapped surfaces and an event horizon are related
by the singularity theorems. 
It would be interesting to explore such a connection.

Finally, we point out the important difference between DTTSs and
trapped surfaces (or apparent horizons).  
On one hand, positions at which trapped surfaces exist
cannot be observed in principle since they 
are formed within an event horizon,
unless cosmic censorship or
the null energy condition is violated.
On the other hand, since DTTSs are formed
and remain outside the event horizon,
positions at which DTTSs exist 
are observable. For this reason,
we expect that the concept of DTTSs would also 
become important in the context of observations of
strong gravity regions in dynamical evolutions.

\ack

H.Y. thanks Hideki Ishihara and Ken-ichi Nakao for helpful comments.
H.Y. is supported by the Grant-in-Aid for
Scientific Research (C) No. JP18K03654 from the Japan Society for
the Promotion of Science (JSPS).
K. I. is supported by JSPS Grant-in-Aid for Young Scientists (B)
No. JP17K14281.
T. S. is supported by Grant-in-Aid for Scientific Research (C) No. JP16K05344
from JSPS.
K.I. and T.S. are
also supported by Scientific Research (A) No. JP17H01091
and  in part by JSPS Bilateral Joint Research Projects
(JSPS-NFR collaboration) ``String Axion Cosmology.''
The work of H.Y. is partly supported by
Osaka City University Advanced Mathematical Institute
(MEXT Joint Usage/Research Center on Mathematics and Theoretical Physics).

\appendix

%
%======================================%
%<<<<<<<<<<<< APPENDIX A  >>>>>>>>>>>>>>%
%======================================%
%
\section{Equations for marginally DTTSs in the Brill--Lindquist initial data}
\label{Appendix-A}

In this appendix we derive the equations
for marginally DTTSs in the Brill--Lindquist initial data 
studied in Sect.~\ref{section4}.

In Eqs.~\eqref{coordinate-transformation-tilder-r-theta}
and \eqref{coordinate-transformation-tildetheta-r-theta},
the coordinates $(r,\theta)$ are introduced.
Transforming the metric from $(\tilde{r}, \tilde{\theta}, \phi)$
coordinates to $(r,\theta,\phi)$ coordinates,
we have the nonzero components 
\begin{subequations}
\begin{eqnarray}
  \varphi^2 &=& \varPsi^4\left[1+(r+h)^2p_{,r}^2\right],
  \\
  \gamma_{r\theta} &=& \varPsi^4\left[h^\prime-(r+h)^2(1-p_{,\theta})p_{,r}\right]
  \\
  h_{\theta\theta} &=& \varPsi^4\left[h^{\prime 2}+(r+h)^2(1-p_{,\theta})^2\right],
  \\
  h_{\phi\phi} &=& \varPsi^4(r+h)^2\sin^2\tilde{\theta},
\end{eqnarray}
\end{subequations}
in the spatial part of the metric in Eq.~\eqref{metric-neighborhood-of-sigma0}.
We determine the function 
$p(r,\theta)$ so that $(r,\theta)$ become orthogonal, i.e. $\gamma_{r\theta}=0$.
On $\sigma_0$ (that is, $r=0$), this means 
\begin{equation}
  \left.p_{,r}\right|_{\sigma_0} = \frac{h^\prime}{h^2},
  \qquad
  \left.p_{,r\theta}\right|_{\sigma_0} =
  \frac{h^{\prime\prime}}{h^2}-\frac{2h^{\prime 2}}{h^3},
\end{equation}
because $p_{,\theta}=p_{,\theta\theta}=0$ holds on $\sigma_0$.
In the coordinates $(r,\theta,\phi)$,
the coordinate components of the extrinsic curvature $k_{ab}$
are calculated by $k_{ij} = h_{ij,r}/2\varphi$ on $\sigma_0$,
and the orthonormal components $k_1$ and $k_2$ are given by
$k_1=k_{\theta\theta}/h_{\theta\theta}$
and $k_2=k_{\phi\phi}/h_{\phi\phi}$. 
The result is
\begin{subequations}
\begin{eqnarray}
k_1&=&-\frac{h}{\varPsi^2(h^2+h^{\prime 2})^{3/2}}
(h^{\prime\prime}+C),
\label{k1_BL}
  \\
  k_2 &=& \frac{D}{\varPsi^2h\sqrt{h^2+h^{\prime 2}}},
  \label{k2_BL}
\end{eqnarray}
\end{subequations}
with
\begin{subequations}
\begin{eqnarray}
  C &=& -h-2\frac{h^{\prime 2}}{h}
  -2\left(\frac{\varPsi_{,\tilde{r}}}{\varPsi}
  -\frac{h^\prime}{h^2}\frac{\varPsi_{,\tilde{\theta}}}{\varPsi}\right)
  \left(h^2+h^{\prime 2}\right),
  \label{C_BL}
  \\
D &=& h-\cot\theta h^\prime +2\left(\frac{\varPsi_{,\tilde{r}}}{\varPsi}
-\frac{h^\prime}{h^2}\frac{\varPsi_{,\tilde{\theta}}}{\varPsi}\right)h^2.
\label{D_BL}
\end{eqnarray}
\end{subequations}
From the induced metric in Eq.~\eqref{induced-metric-sigma0-BL} on $\sigma_0$,
the Ricci scalar ${}^{(2)}R$ is calculated as
\begin{equation}
  \frac12{}^{(2)}R = \frac{1}{\varPsi^4(h^2+h^{\prime 2})}
  \left(Ah^{\prime\prime}+B\right),
\end{equation}
with 
\begin{subequations}
\begin{equation}
  A=-\frac{2\varPsi_{,\tilde{r}}}{\varPsi}
  + \frac{1}{h^2+h^{\prime 2}}\left(2\frac{\varPsi_{,\tilde{r}}h^\prime+\varPsi_{,\tilde{\theta}}}{\varPsi}h^\prime-h\right)
  +\frac{h^\prime}{h^2+h^{\prime 2}}\cot\theta,
\end{equation}
\begin{multline}
  B=1-\frac{2}{\varPsi}\left[\varPsi_{,\tilde{r}\tilde{r}}h^{\prime 2}+2\varPsi_{,\tilde{r}\tilde{\theta}}h^\prime
    +\varPsi_{,\tilde{\theta}\tilde{\theta}}-\frac{(\varPsi_{,\tilde{r}}h^\prime+\varPsi_{,\tilde{\theta}})^2}{\varPsi}\right]
  \\
  -\frac{h^{\prime 2}}{h(h^2+h^{\prime 2})}
  \left(2\frac{\varPsi_{,\tilde{r}}h^\prime+\varPsi_{,\tilde{\theta}}}{\varPsi}h^\prime-h\right)
  -\left[\frac{h^\prime}{h^2+h^{\prime 2}}\left(h+2\frac{h^{\prime 2}}{h}\right)
    +2\frac{\varPsi_{,\tilde{r}}h^\prime+\varPsi_{,\tilde{\theta}}}{\varPsi}\right]\cot\theta.
\end{multline}
\end{subequations}
Below, we study the equation for a marginally DTTS,
Eq.~\eqref{Equation-for-a-marginally-DTTS-time-symmetric}, 
for the cases $k_1\le k_2$ and $k_1\ge k_2$, separately.

\subsection{The case $k_1\le k_2$}

In the case $k_1\le k_2$, we put $\mathrm{max}(k_1, k_2) = k_2$
in Eq.~\eqref{Equation-for-a-marginally-DTTS-time-symmetric},
which reduces to
\begin{equation}
    h^{\prime\prime} = \frac{-2CD+(D^2/h^2-B)(h^2+h^{\prime 2})}{2D+A(h^2+h^{\prime 2})}.
  \label{equation-prolate-case}
\end{equation}
in the range $0<\theta<\pi$.
Since Eq.~\eqref{equation-prolate-case} includes $\cot\theta$,
we have to regularize the equation at the poles $\theta=0$ and $\pi$.
Since $h^\prime=0$ and $\varPsi_{,\tilde{\theta}}=0$
holds at the poles
for axisymmetric initial data and an axisymmetric surface,
the terms including $\cot{\tilde{\theta}}$ behave 
as $h^\prime\cot\theta\to h^{\prime\prime}$
and $\varPsi_{,\tilde{\theta}}\cot\theta\to \varPsi_{,\tilde{\theta}\tilde{\theta}}$
in the limit $\theta\to 0$ and $\pi$.
Then, a quadratic equation for $h^{\prime\prime}$ is derived as
\begin{equation}
  2h^{\prime\prime 2}
  +4\tilde{C}h^{\prime\prime}
  + 3\tilde{C}^2
  =h^2\left(1-\frac{4\varPsi_{,\tilde{\theta}\tilde{\theta}}}{\varPsi}\right),
\end{equation}
where $\tilde{C}=-(2h^2\varPsi_{,\tilde{r}}/\varPsi+h)$
is the value of $C$ at the poles. 
Then, a solution with double sign is obtained for $h^{\prime\prime}$, 
and we must choose a
physically appropriate sign.
This can be done by considering the
spherically symmetric case $z_0=0$, because 
$h=(1+\sqrt{3}/2)GM$ is a solution and
the sign must be chosen so that $h^{\prime\prime}=0$
is realized. In this way, we obtain 
\begin{equation}
  h^{\prime\prime} =h+2\frac{\varPsi_{,\tilde{r}}}{\varPsi}h^2
  - \frac{\sqrt{2}h}{\varPsi}
  \sqrt{-\left[\varPsi\varPsi_{,\tilde{\theta}\tilde{\theta}}
      +\varPsi_{,\tilde{r}}h(\varPsi+\varPsi_{,\tilde{r}}h)\right]},
  \label{equation-regularized-at-poles}
\end{equation} 
for $\theta=0$ and $\pi$.

\subsection{The case $k_1\ge k_2$}
In the case $k_1\ge k_2$, we set $\mathrm{max}(k_1, k_2) = k_1$
in Eq.~\eqref{Equation-for-a-marginally-DTTS-time-symmetric}.
Then, the equation is reduced to
\begin{equation}
  h^{\prime\prime 2}
  + 2\left[
    C-\frac{h^2+h^{\prime 2}}{h^2}D-\frac{(h^2+h^{\prime 2})^2}{2h^2}A
    \right]h^{\prime\prime}
  +C^2-2\frac{h^2+h^{\prime 2}}{h^2}CD-\frac{(h^2+h^{\prime 2})^2}{h^2}B=0.
  \label{equation-oblate-case-original}
\end{equation}
Solving this equation with respect to $h^{\prime\prime}$,
a solution with double sign is obtained. An appropriate sign is
chosen by requiring that $h^{\prime\prime}=0$
is realized in the spherically symmetric case. The result is
\begin{equation}
  h^{\prime\prime} =
  -C + \frac{h^2+h^{\prime 2}}{h^2}\left[D+\frac12(h^2+h^{\prime 2})A\right]
  - \frac{h^2+h^{\prime 2}}{h^2}\sqrt{\left[D+\frac12(h^2+h^{\prime 2})A\right]^2+h^2(B-AC)},
  \label{equation-oblate-case}
\end{equation}
for $0<\theta<\pi$. At the poles $\theta=0$ and $\pi$, the
regularized equation is reduced to the same equation
as the case $k_1\le k_2$, Eq.~\eqref{equation-regularized-at-poles},
because $k_1=k_2$ is satisfied at the poles
by regular surfaces.

%
%======================================%
%<<<<<<<<<<<< APPENDIX B  >>>>>>>>>>>>>>%
%======================================%
%
\section{Equations for marginally DTTSs in the Majumdar--Papapetrou spacetime}
\label{Appendix-B}

The equations for marginally DTTSs in the Majumdar--Papapetrou spacetime
are obtained by the same basic procedure as
the Brill--Lindquist case. We span the 
spherical-polar coordinates $(\tilde{r}, \tilde{\theta}, \phi)$, 
parametrize the surface $\sigma_0$ as $\tilde{r}=h(\tilde{\theta})$,
and introduce the coordinates $(r,\theta,\phi)$ 
in the same manner as the Brill--Lindquist case.
The geometrical quantities on $t=\mathrm{const.}$
are obtained by replacing $\varPsi$ with $U^{1/2}$
in the Brill--Lindquist cases given in Appendix~\ref{Appendix-A}. 
Then, the formulas for $k_1$ and $k_2$
are
\begin{subequations}
\begin{eqnarray}
k_1&=&-\frac{h}{U(h^2+h^{\prime 2})^{3/2}}
(h^{\prime\prime}+C),
\label{k1_MP}
  \\
  k_2 &=& \frac{D}{Uh\sqrt{h^2+h^{\prime 2}}},
\label{k2_MP}
\end{eqnarray}
\end{subequations}
with
\begin{subequations}
\begin{eqnarray}
  C &=& -h-2\frac{h^{\prime 2}}{h}
  -\left(\frac{U_{,\tilde{r}}}{U}
  -\frac{h^\prime}{h^2}\frac{U_{,\tilde{\theta}}}{U}\right)
  \left(h^2+h^{\prime 2}\right),
  \label{C_MP}
  \\
D &=& h-\cot\theta h^\prime +\left(\frac{U_{,\tilde{r}}}{U}
-\frac{h^\prime}{h^2}\frac{U_{,\tilde{\theta}}}{U}\right)h^2,
\label{D_MP}
\end{eqnarray}
\end{subequations}
and the formula for ${}^{(2)}R$ is
\begin{equation}
  \frac12{}^{(2)}R = \frac{1}{U^2(h^2+h^{\prime 2})}
  \left(Ah^{\prime\prime}+B_1\right),
\end{equation}
with
\begin{subequations}
\begin{equation}
  A=-\frac{U_{,\tilde{r}}}{U}
  + \frac{1}{h^2+h^{\prime 2}}\left(\frac{U_{,\tilde{r}}h^\prime+U_{,\tilde{\theta}}}{U}h^\prime-h\right)
  +\frac{h^\prime}{h^2+h^{\prime 2}}\cot\theta,
\end{equation}
\begin{multline}
  B_1=1-\frac{1}{U}\left[U_{,\tilde{r}\tilde{r}}h^{\prime 2}+2U_{,\tilde{r}\tilde{\theta}}h^\prime
    +U_{,\tilde{\theta}\tilde{\theta}}-\frac{(U_{,\tilde{r}}h^\prime+U_{,\tilde{\theta}})^2}{U}\right]
  \\
  -\frac{h^{\prime 2}}{h(h^2+h^{\prime 2})}
  \left(\frac{U_{,\tilde{r}}h^\prime+U_{,\tilde{\theta}}}{U}h^\prime-h\right)
  -\left[\frac{h^\prime}{h^2+h^{\prime 2}}\left(h+2\frac{h^{\prime 2}}{h}\right)
    +\frac{U_{,\tilde{r}}h^\prime+U_{,\tilde{\theta}}}{U}\right]\cot\theta.
\end{multline}
\end{subequations}
The important difference is that there is a nonzero contribution
from the radial pressure $P_r$. 
The energy-momentum tensor of electromagnetic fields is given by
\begin{equation}
T_{ab} = \frac{1}{4\pi}\left({F_{a}}^{c}F_{bc}-\frac14g_{ab}F_{cd}F^{cd}\right),
\end{equation}
with the electromagnetic tensor $F_{ab}=\nabla_aA_b-\nabla_bA_a$. 
The spatial components of the energy-momentum tensor,
$S_{ab}={q_{a}}^{c}{q_{b}}^{d}T_{cd}$, are calculated as
\begin{equation}
S_{ab}=    \frac{1}{4\pi G}\left[
    -\frac{D_aU D_bU}{U^2}
    +\frac12q_{ab}\frac{(DU)^2}{U^2}\right]
  \label{MP-energymomentum-spatial}
\end{equation}
from Eq.~\eqref{MP:electromaginetic-potential}, and thus we have
\begin{equation}
  8\pi G P_r 
  =
  \frac{(D_aU)(D^aU)}{U^2}-2\frac{(\hat{r}^aD_aU)^2}{U^2}
  =\frac{B_2}{U^2(h^2+h^{\prime 2})},
\end{equation}
with 
\begin{equation}
  B_2=
  (h^2+h^{\prime 2})\left(\frac{U_{,\tilde{r}}^2}{U^2}+\frac{U_{,\tilde{\theta}}^2}{h^2U^2}\right)
    -2h^2\left(\frac{U_{,\tilde{r}}}{U}-\frac{h^\prime}{h^2}\frac{U_{,\tilde{\theta}}}{U}\right)^2,
\end{equation}
where we used the fact that
the components of $\hat{r}^a$ in the $(t, \tilde{r}, \tilde{\theta}, \phi)$ coordinates 
are given by
\begin{equation}
\hat{r}^{\mu} = \frac{1}{U\sqrt{1+h^{\prime 2}/h^2}}
\left(0,1,-h^{\prime}/h^2,0\right)
\end{equation}
on $\sigma_0$.
Defining
\begin{equation}
B=B_1+B_2, 
\end{equation}
the equation for a marginally DTTS is given by
the same form as Eq.~\eqref{equation-prolate-case}
in the case $k_1\le k_2$, and by the same form as Eq.~\eqref{equation-oblate-case}
in the case $k_1\ge k_2$, in the range $0<\theta<\pi$.
At the poles, the equation is regularized as
\begin{equation}
  h^{\prime\prime}
  =
  \frac{h}{U}
  \left[U+hU_{,\tilde{r}}-\sqrt{-\left[UU_{,\tilde{\theta}\tilde{\theta}} + hU_{,\tilde{r}}(U+hU_{,\tilde{r}})\right]}\right].
\end{equation}

\end{document}